\documentclass[twocolumn]{aastex631}

\usepackage{gensymb} 
\usepackage[T1]{fontenc}
\usepackage{tgbonum}
\usepackage{tablefootnote} 
\usepackage{xcolor}
\usepackage{bm}

\begin{document}

\title{A Spectroastrometric Study of the Low-velocity Wind from DG Tau A\footnote{This research is based on data collected at the Subaru Telescope, which is operated by the National Astronomical Observatory of Japan. We are honored and grateful for the opportunity to observe the Universe from Maunakea, which has cultural, historical, and natural significance in Hawaii.}}

\correspondingauthor{Yu-Ru Chou}
\author[0009-0004-9608-6132]{Yu-Ru Chou}
\email{yrchou@mpe.mpg.de}
\affiliation{Institute of Astronomy and Astrophysics, Academia Sinica, 11F of Astronomy-Mathematics Building, No.1, Sec. 4, Roosevelt Rd, Taipei 10617, Taiwan, R.O.C.}
\affiliation{Institute of Astronomy and Department of Physics, National Tsing Hua University, Hsinchu 30013, Taiwan; Institute of Astronomy and Astrophysics, Academia Sinica, No.1, Sec. 4, Roosevelt Road, Taipei 10617, Taiwan}
\affiliation{Max-Planck-Institut f{\"u}r extraterrestrische Physik, Giessenbachstrasse 1, D-85748 Garching, Germany}

\author[0000-0001-9248-7546]{Michihiro Takami}
\affiliation{Institute of Astronomy and Astrophysics, Academia Sinica, 11F of Astronomy-Mathematics Building, No.1, Sec. 4, Roosevelt Rd, Taipei 10617, Taiwan, R.O.C.}

\author[0000-0001-5522-486X]{Shin-Ping Lai}
\affiliation{Institute of Astronomy and Department of Physics, National Tsing Hua University, Hsinchu 30013, Taiwan; Institute of Astronomy and Astrophysics, Academia Sinica, No.1, Sec. 4, Roosevelt Road, Taipei 10617, Taiwan}

\author[0000-0002-3741-9353]{Emma Whelan}
\affiliation{Maynooth University, Department of Physics, National University of Ireland Maynooth, Maynooth, Co. Kildare, Ireland}

\author[0000-0002-2530-4137]{Noah B. Otten}
\affiliation{Maynooth University, Department of Physics, National University of Ireland Maynooth, Maynooth, Co. Kildare, Ireland}
\affiliation{European Southern Observatory, Alonso de Córdova 3107, Vitacura, Casilla 19001, Santiago de Chile, Chile}

\author[0000-0001-8060-1321]{Min Fang}
\affiliation{Purple Mountain Observatory, Chinese Academy of Sciences, Nanjing, China; University of Science and Technology of China, Hefei, China}

\author[0000-0001-8813-9338]{Akito Tajitsu}
\affiliation{Subaru Telescope Okayama Branch Office, National Astronomical Observatory of Japan, 3037-5, Honjou, Kamogata, Asakuchi, Okayama 719-0232, Japan}

\author[0000-0001-7076-0310]{Masaaki Otsuka}
\affiliation{Okayama Observatory, Kyoto University, Kamogata, Asakuchi, Okayama 719-0232, Japan}

\author[0000-0001-8385-9838]{Hsien Shang}
\affiliation{Institute of Astronomy and Astrophysics, Academia Sinica, 11F of Astronomy-Mathematics Building, No.1, Sec. 4, Roosevelt Rd, Taipei 10617, Taiwan, R.O.C.}

\author[0000-0002-1624-6545]{Chun-Fan Liu}
\affiliation{Institute of Astronomy and Astrophysics, Academia Sinica, 11F of Astronomy-Mathematics Building, No.1, Sec. 4, Roosevelt Rd, Taipei 10617, Taiwan, R.O.C.}

\author[0000-0003-3095-4772]{Jennifer Karr}
\affiliation{Institute of Astronomy and Astrophysics, Academia Sinica, 11F of Astronomy-Mathematics Building, No.1, Sec. 4, Roosevelt Rd, Taipei 10617, Taiwan, R.O.C.}

\author[0000-0003-0376-6127]{Aisling Murphy}
\affiliation{Institute of Astronomy and Astrophysics, Academia Sinica, 11F of Astronomy-Mathematics Building, No.1, Sec. 4, Roosevelt Rd, Taipei 10617, Taiwan, R.O.C.}

\begin{abstract}
We obtained high spectral resolution spectra ($\Delta v$ $\sim$ 2.5 km s$^{-1}$) for DG Tau A
from 4800 Å to 7500 Å using Subaru High Dispersion Spectrograph (HDS) for the first time.
The low-velocity components (LVCs, |$v$| < 100 km s$^{-1}$) 
were observed in the [O I] 5577, 6300, 6364 Å, [S II] 6716, 6731 Å lines. 
%
{The offset position spectra observed in a component within the LVC velocity range between --16 km s$^{-1}$ to --41 km s$^{-1}$, namely, LVC-M, show a ``negative velocity gradient'', supporting the presence of a wide-angled wind.}
With 12-70 au wind lengths measured using spectroastrometry, we estimate a lower limit to the wind mass-loss rate of $\sim$10$^{-8}$ M$_\odot$ yr$^{-1}$.
In addition to the LVCs, we identify two high-velocity components (HVCs, |$v$| > 100 km s$^{-1}$) associated with the collimated jet in 26 lines ([N I], [N II], [O I], [O II], [O III], [S II], [Ca II], [Fe II], H$\alpha$, H$\beta$, He I).
The one 
with a clear spatial offset from the star
($n_e$ $\sim$10$^4$ cm$^{-3}$, HVC1) is associated with an internal shock surface of the jet, while the other at the base ($n_e$ $\sim$10$^6$ cm$^{-3}$, HVC2) may be a stationary shock component.
We find that the observed line profiles and the spatial scales of the LVC emission do not agree with the existing predictions for photoevaporative or magnetohydrodynamical (MHD) disk winds. These could be explained by the X-wind model, but synthetic observations are required for detailed comparisons.
\end{abstract}

\accepted{by the Astrophysical Journal, March 1, 2025}


\section{Introduction} \label{sec:intro}

Many T-Tauri stars are associated with high-velocity collimated jets (typically $v \gtrsim 50$ km s$^{-1}$) and low-velocity winds (typically $v < 50$ km s$^{-1}$), that potentially contribute to the dispersal of protoplanetary disks and may therefore affect planet formation and migration
\citep[e.g.,][]{Alexander14, Ercolano23}.
Understanding the connection between wind and jet mass loss rates requires high-spectral and high-angular resolution estimated from many young sources.
Recent studies have suggested that the mass-loss rate of the wind is much higher than that of the jet \citep[e.g., $\sim$50 times for the HH 30 protostar;][]{Louvet18,Bacciotti99_HH30}
and may even be comparable to the mass accretion rate \citep{Fang18, Louvet18, deValon20} \footnote{See also \citet{Pascucci23} for a review of similar studies of Class 0 and I protostars.}.
If this is the case, the wind mass loss directly affects the mass evolution of the disk in which planets are forming.
However, the wind mass-loss rates measured to date have suffered from uncertainties in physical parameters such as the
gas temperature and the wind spatial extension \citep{Fang18}.
In addition,
line emission from the wind
may include surrounding entrained gas not ejected from the star-disk system \citep{Louvet18}.
In summary, the low-velocity winds may hold a key to understanding the mass evolution of the disks and associated planet formation, however, their importance is not yet clear.

The optical forbidden line emission from T Tauri stars is known to trace their outflows \citep{Hamann94, Hartigan95, Hirth97}. 
These lines usually only show blueshifted emission,
with the absence of the redshifted emission explained by obscuration due to the optically thick circumstellar disk.
The line profiles often exhibit two velocity components: a high-velocity component (HVC, 
--50
to --200 km s$^{-1}$) associated with a collimated jet, and a distinct low-velocity component (LVC, --5
to --20 km s$^{-1}$).
The LVC is expected to trace 
wide-angled
winds
distinct from the HVC, because (1) its line flux ratios imply an electron density higher than the HVC; and (2) its radial velocity is comparable to the linewidth \citep[][]{Hartigan95,Eisloffel00_PPIV}.
Further studies found that  the LVC line profiles associated with many T-Tauri stars can be decomposed into two Gaussian
components
with similar peak velocities, the broad (BC) and narrow components (NC), with typical full width half maximum (FWHM) velocities of approximately 100 and 30 km s$^{-1}$, respectively \citep[e.g., ][]{Rigliaco13,Nisini24}.

There are three proposed origins for the LVCs:
magnetohydrodynamical (MHD)
wind, a photoevaporative wind 
(PE wind), and 
gravitationally bound gas at or above the disk surface
\citep{Rigliaco13,Simon16,Banzatti19,Weber20}.
%
The X-wind model is included within the category of MHD winds.
The X-wind originates from the region where stellar magnetic field lines connect with the protoplanetary disk near the 
inner disk edge. These winds are characterized by high velocities, typically around 150 km s$^{-1}$, and have a wide angular distribution as they are launched from the inner disk \citep{Shu00}.
On the other hand, the classical
MHD disk winds
would be launched from 
a wide range of disk radii, from near the gas co-rotation radius ($\sim$0.1 au) through more extended disk radii,
either due to magnetocentrifugal force \citep[e.g.,][]{Blandford82,Konigl00,Ogilvie12} or magnetic pressure \citep[e.g.,][]{Uchida85,Bai17}.

The PE wind would occur when thermal energy from high-energy radiation, such as extreme ultraviolet (EUV; 13.6--100 eV), far ultraviolet (FUV; 6--13.6 eV), and X-ray (0.1--10 keV) radiation heats the disk surface, 
allowing gas from the disk to escape
\citep{Alexander14}.
The EUV, FUV, and X-ray photons would penetrate and heat the gas at column densities up to $\sim$$10^{20}$, $\sim$$10^{24}$, and 10$^{20-22}$ cm$^{-2}$, respectively \citep{Pascucci23}.
A detailed investigation of these winds is needed to understand from what region and to what extent the low-velocity winds remove gas mass and angular momentum and, therefore,
how it affects disk evolution.

One of the main observational challenges
for probing the origin of the outflows is the limit of high spatial resolution \citep[e.g.,][]{Eisloffel00_PPIV}.
In the lack of high spatial resolution, the spectro-astrometry (SA) technique is used to observe the relative position of different line centroids
 of the point-spread-function (PSF) on scales of a few milliarcseconds (mas) \citep[e.g.,][]{Bailey98b, Bailey98a, Whelan08}.
This technique has been used to investigate emission line displacements relative to the stellar position \citep{Takami01, Whelan21}. 
\citet{Takami01} used SA to reveal that the redshifted H$\alpha$ outflow from RU Lupi could indicate a disk gap at 3-4 au. 
\citet{Whelan21} applied SA to forbidden line emission associated with RU Lupi and AS 205 with a significantly better spectral resolution ($\Delta v =$ 7 km s$^{-1}$).
These authors demonstrated that the position spectra for the RU Lupi LVC-NC show the negative velocity gradient expected for an MHD disk wind (See Section \ref{sec: major} for details).\\

Our target star, DG Tau A, is a T Tauri star with strong outflows 
(jets/winds) in atomic and ionic emission lines at optical and infrared wavelengths \citep[e.g.,][]{Lavalley00,Bacciotti00,Pyo03,Davis03,White14a,LiuC16,Takami23}.
DG Tau A, at a distance of 138 pc \citep{GaiaDR3}, is located in the Taurus molecular cloud.
The observed flat spectral energy distribution (SED) in the infrared suggests that DG Tau A is transitioning from a Class I protostar to a Class II classical T Tauri star (CTTS) \citep{Adams90, Calvet94}.
Several optical high-resolution spectroscopic observations have shown evidence
for variable forbidden line emission from LVC-BC (at a radial centroid velocity of $\sim$ --30 km s$^{-1}$ from the star), LVC-NC (--4 to --20 km s$^{-1}$), and HVC (--150 to --420 km s$^{-1}$) \citep[e.g.,][; Otten et al., in prep.]{Hartigan95,Simon16,Banzatti19,Giannini19,Nisini24,Pyo24,LiuC16}. In addition to these components, \citet{Giannini19} identified a medium-velocity component (MVC) centered at $\sim$ --60 km s$^{-1}$. 


In this paper, we present our analysis of the observations of this star taken with the High-Dispersion Spectrograph (HDS) on the Subaru 8.2-m telescope. A combination of high signal-to-noise and wide wavelength coverage (4800--7500 Å) allowed us to observe a number of forbidden emission lines associated with outflowing gas.  The high signal-to-noise also allowed us to apply the SA technique to the major emission lines in order to investigate the mass accretion rate of the 
LVC wind(s)
more accurately than before.
Furthermore, the spectral resolution of $\Delta v \simeq$ 2.5 km s$^{-1}$ for our observations was approximately three times higher than those of \citet{Banzatti19} executed using the High Resolution Echelle Spectrometer (HIRES) on the 10-m Keck I telescope and the Magellan Inamori Kyocera Echelle (MIKE) spectrograph on the 6.5-m Magellan telescope; and comparable to those of \citet{Giannini19} and \citet{Nisini24} executed using GIARPS at the 3.6-m Galileo National Telescope (TNG) telescope. Our observations allowed us to identify a number of emission components, some of which have not previously been reported.
The major goal of this study is to demonstrate how these improved observations can be used to investigate the nature of the LVCs.

The rest of the paper is organized as follows.
In Section \ref{sec: method}, we describe the observations, emission line selection, and data reduction. In Section \ref{sec: results}, we present the kinematic structures and physical conditions observed in a variety of emission lines. In Section \ref{sec: discus}, we discuss the possible nature of the individual velocity components for the jet and winds we have identified. 
In Section \ref{sec: mechanism}, we compare our results with various wind models.
We summarize our findings and give conclusions in Section \ref{sec: summary}.\\

\section{Observations and Data Reduction} \label{sec: method}
Observations were carried out on December 6 and 7, 2003, with Subaru/HDS. The echelle grating mode with a 3" $\times$ 0.3" slit provides a high spectral resolution of $\sim$ 2.5 km s$^{-1}$ with an observable wavelength range of 4800 Å to 7500 Å. The pixel scale of 0".138 ensures adequate sampling of the seeing profile (FWHM $\sim$ 0".6-1".2 during our observations), enabling accurate spectroastrometric measurements \citep{Bailey98b, Bailey98a}.
Spectra were obtained at four slit position angles (PAs) with respect to the jet PA of 224\arcdeg~
measured by previous imaging and spectro-imaging \citep[e.g.,][]{Kepner93, Lavalley97, Takami23}. 
Note that 0\arcdeg~is north and the PA increases counterclockwise.
The four PAs are 224\degree~and 44\degree, which are aligned with the jet, and 134\arcdeg~and 314\arcdeg, which are perpendicular to the jet. The spectra at anti-parallel slit angles were used for the SA measurements as described below.
The spectra at each PA were obtained with two exposures (900 s$\times$2 for PA=224\degree~and 44\degree, respectively; 600 s$\times$2 for PA=134\degree~and 314\degree, respectively). 
HD 42784 (a B8V star) was observed at an air mass of 1.26 with an exposure time of 180 s for telluric correction and flux calibration. The telluric correction was made by scaling the absorption features in the telluric standard to the spectra for the target star observed at air masses of 1.0-2.05.
VA68 (K4), a main-sequence star with a spectral type similar to DG Tau A (K6), was observed with an exposure time of 600 s to remove the stellar absorption from the spectra of the target star.\\

The data were reduced using the pipeline developed by a group member for HDS. The pipeline is based on the Image Reduction and Analysis Facility (IRAF), and it includes subtraction of bias (using the overscan region of the detector); removals of bad pixels, scattered light and crosstalks; flat fielding, extraction of the individual echelle orders, and wavelength calibration for the echelle data. Telluric and stellar absorption were removed using the Python numpy and scipy packages. We corrected for a measured heliocentric velocity of DG Tau A of 19 km s$^{-1}$. Finally, we combined the 2-D spectra of PAs of 224\degree~and 44\degree~to obtain the position-velocity (PV) diagrams for the emission along the jet, and the PAs of 314\degree~and 134\degree~to obtain PV diagrams perpendicular to the jet.\\

We applied the SA technique by fitting a Gaussian for each wavelength along the spatial axis of the 2-D spectrum, thus producing a position spectrum. Any instrumental effects in the position spectra were accurately eliminated by subtracting those with opposite position angles (224\degree---44\degree~or 314\degree---134\degree). 
When we combined the spectra, we resampled the spectra to a 1 km s$^{-1}$ grid.
After this subtraction, the position spectra were centered at the continuum position.
To determine the true position of the line emission, contamination by the continuum was removed using the following equation \citep[e.g.,][]{Takami01,Whelan08}:
\begin{equation} \label{eq:scaling_position_spectra}
x_\mathrm{true} = \frac{F_\mathrm{cont}(\lambda)+F_\mathrm{line}(\lambda)}{F_\mathrm{line}(\lambda)}x_\mathrm{obs},
\end{equation}
where $x_\mathrm{true}$ is the actual offset for the emission line; $x_\mathrm{obs}$ is the measured offset before applying this equation; and $F_\mathrm{cont}(\lambda)$ and $F_\mathrm{line}(\lambda)$ are the continuum and line fluxes, respectively.
We note that the SA measurements are reliable for line emission within approximately half the seeing size of the stellar position. For our data, this is within $\sim$0\farcs5, i.e., the cases for which centroiding using a Gaussian is reasonably reliable.\\

The signal-to-noise of the SA measurements is determined from the photon noise in our observations. Most emission lines presented in this paper are significantly fainter than the continuum, therefore the photon noise is primarily determined by the continuum. Using the position spectra before applying Equation (\ref{eq:scaling_position_spectra}), we measured a typical positional accuracy of 4 mas for [O I] 6300 Å. The noise level at each velocity and offset values are scaled when we apply Equation (\ref{eq:scaling_position_spectra}). Our brightest target lines have fluxes comparable to those of the continuum, and as a result, the noise level for these lines is slightly larger, by 30 to 50 \%.

In our search for emission lines from the outflows, we
visually inspected the continuum-subtracted 2D spectra and searched for extended emission over the entire wavelength coverage.
Furthermore, we carefully inspected the spectra at
the 
forbidden line wavelengths
listed in Table 5 of \citet{Hamann94}. The detected lines and their rest wavelengths, Einstein A coefficients, and upper-level energies are listed in Table \ref{tab: line}.\\

\begin{longrotatetable}
\begin{deluxetable*}{llllllllll}
\tablewidth{5pt} 
\tablecaption{Properties of the detected lines \label{tab: line}
}
\tablehead{
\multicolumn{2}{c}{Transition}  &
$\lambda$\tablenotemark{\tiny b}  & 
$A$\tablenotemark{\tiny b}  & 
$E_u$\tablenotemark{\tiny b}&
$n_\mathrm{crit}$\tablenotemark{\tiny c}  & $I_\mathrm{HVC1}$ \tablenotemark{\tiny d} & $I_\mathrm{HVC2}$ \tablenotemark{\tiny e} & $v_\mathrm{p,HVC1}$\tablenotemark{\tiny f} & $v_\mathrm{p,HVC2}$\tablenotemark{\tiny f}\\
& &(Å) & (s$^{-1}$)&(cm$^{-1}$)&(cm$^{-3}$) & ($10^{-18}$ W m$^{-2}$ arcsec$^{-1}$ Å$^{-1}$)& ($10^{-18}$ W m$^{-2}$ arcsec$^{-1}$ Å$^{-1}$) & (km s$^{-1}$) & (km s$^{-1}$)
}
\startdata 
        [N I] & 5200\tablenotemark{\tiny a} & 5200.257 & { 7.56$\times$10$^{-6}$} & 19,224.46 & { 1.5$\times$10$^3$} & < 0.3 & 8 $\pm$ 1 &-\tablenotemark{\tiny g} &--160 \\ 
        {[N II]} & 6583 & 6583.450 & { 2.92$\times$10$^{-3}$} & 15,316.20 & { 8.4$\times$10$^4$} & 9.3 $\pm$ 0.2 & 7.1 $\pm$ 0.5 & --251 & --184\\ 
        ~ & 5755 & 5754.590 & 1.14 & 32,688.80 & { 1.7$\times$10$^7$} & 0.62 $\pm$ 0.09 & 1.9 $\pm$ 0.2 & -\tablenotemark{\tiny g} & --207\\ 
        {[O I]} & 6300 & 6300.304 & { 5.65$\times$10$^{-3}$} & 15,867.86 & { 1.6$\times$10$^6$} & 13.50 $\pm$ 0.09 & 56.1 $\pm$ 0.4 & --248 & --181\\ 
        ~ & 6364 & 6363.780 & { 1.82$\times$10$^{-3}$} & 15,867.86 & { 1.6$\times$10$^6$} & 4.61 $\pm$ 0.08 & 18.8 $\pm$ 0.2 & --249 & --182 \\ 
        ~ & 5577 & 5577.340 & 1.26 & 33,792.58 & { 1.3$\times$10$^8$} & 0.26 $\pm$ 0.06 & 3.1 $\pm$ 0.4 & -\tablenotemark{\tiny g} & --200\\ 
        {[O II]} & 7330 & 7329.68 & { 9.32$\times$10$^{-3}$} & 40,470.00 & { 2.9$\times$10$^6$} & -\tablenotemark{\tiny h} & 8.4 $\pm$ 0.8 & -\tablenotemark{\tiny h} &--184\\ 
        {[O III]} & 5007 & 5006.843 & { 1.81$\times$10$^{-2}$} & 20,273.27 & { 6.8$\times$10$^5$} & 1.5 $\pm$ 0.1 & -\tablenotemark{\tiny i} & --239 & -\tablenotemark{\tiny i} \\ 
        ~ & 4959 & 4958.911 & { 6.21$\times$10$^{-3}$} & 20,273.27 & { 6.8$\times$10$^5$} & 0.5 $\pm$ 0.1 & -\tablenotemark{\tiny i} &-\tablenotemark{\tiny g}&-\tablenotemark{\tiny i}\\ 
        {[S II]} & 6731 & 6730.815 & { 6.84$\times$10$^{-4}$} & 14,852.94 & { 6.0$\times$10$^3$} & 4.6 $\pm$ 0.1& 5.4 $\pm$ 0.2 & --248 & -\tablenotemark{\tiny g}\\ 
        ~ & 6716 & 6716.440 & { 2.02$\times$10$^{-4}$} & 14,884.73 & { 4.2$\times$10$^3$} & 2.18 $\pm$ 0.07 & 2.2 $\pm$ 0.1 & --248 &-\tablenotemark{\tiny g}\\ 
        {[Ca II]} & 7291 & 7291.470 & { 1.30} & 13,710.88 & { 4.1$\times$10$^6$} & 8.6 $\pm$ 0.1 & -\tablenotemark{\tiny j} & --254 &-\tablenotemark{\tiny j}\\ 
        {[Fe II]} & 4890 & 4889.623 & { 4.17$\times$10$^{-1}$} & 20,830.55 & { 1.0$\times$10$^7$} & 0.3 $\pm$ 0.1 & -\tablenotemark{\tiny j} &-\tablenotemark{\tiny g}&-\tablenotemark{\tiny j}\\
        ~ & 5159 & 5158.792 & { 4.76$\times$10$^{-1}$} & 21,251.61 & { 1.3$\times$10$^7$} & 1.8 $\pm$ 0.1 & 6.3 $\pm$ 0.3 &-\tablenotemark{\tiny g}&--202\\
        ~ & 5220 & 5220.059 & { 1.10$\times$10$^{-1}$} & 21,581.64 & { 1.2$\times$10$^7$} & 0.37 $\pm$ 0.09 & -\tablenotemark{\tiny i} &-\tablenotemark{\tiny g}&-\tablenotemark{\tiny i}\\ 
        ~ & 5262 & 5261.633 & { 3.39$\times$10$^{-1}$} & 21,430.36 & { 1.3$\times$10$^7$} & 0.9 $\pm$ 0.1 & -\tablenotemark{\tiny j} &-\tablenotemark{\tiny g} &-\tablenotemark{\tiny j}\\ 
        ~ & 5273 & 5273.363 & { 4.72$\times$10$^{-1}$} & 20,830.55 & { 1.0$\times$10$^7$} & < 0.3 & -\tablenotemark{\tiny j} &-\tablenotemark{\tiny g}&-\tablenotemark{\tiny j}\\ 
        ~ & 7155 & 7155.174 & { 1.90$\times$10$^{-1}$} & 15,844.65 & { 2.8$\times$10$^6$}  & 4.50 $\pm$ 0.09 & 13.0 $\pm$ 0.5 & --251 & --195\\ 
        ~ & 7172 & 7171.999 & { 6.02$\times$10$^{-2}$} & 16,369.41 & { 2.7$\times$10$^6$} & 1.33 $\pm$ 0.09 & 4.2 $\pm$ 0.7 & --247 & --192\\
        ~ & 7388 & 7388.167 & { 4.47$\times$10$^{-2}$} & 16,369.41 & { 2.7$\times$10$^6$} & 1.0 $\pm$ 0.1 & 2.5 $\pm$ 0.2 &-\tablenotemark{\tiny g}&-\tablenotemark{\tiny g}\\ 
        ~ & 7453 & 7452.561 & { 6.14$\times$10$^{-2}$} & 15,844.65 & { 2.8$\times$10$^6$} & 1.5 $\pm$ 0.1 & (5.1 $\pm$ 0.4)\tablenotemark{\tiny j} & --249 &-\tablenotemark{\tiny j}\\ 
        {[Fe III]} & 5527 & 5527.026 & { 1.64$\times$10$^2$} & 138,914.09 & >{ 1$\times$10$^{12}$} & 0.61 $\pm$ 0.05 & 2.4 $\pm$ 0.1 & -\tablenotemark{\tiny g} & --179\\ 
        {[Ni II]} & 7378 & 7377.830 & { 2.30$\times$10$^{-1}$} & 13,550.39 & { 1.8$\times$10$^6$} & 2.56 $\pm$ 0.09 & 5.1 $\pm$ 0.8 & --251 & --178\\
        {H$\alpha$} & 6563 & 6562.852 & { 7.96$\times$10$^7$} & 97,492.32 & -\tablenotemark{\tiny k} & 31.6 $\pm$ 0.1 & -\tablenotemark{\tiny l} & --248 & -\tablenotemark{\tiny l} \\
        {H$\beta$} & 4861 & 4861.35 & { 2.90$\times$10$^7$} & 102,823.89 & -\tablenotemark{\tiny k} & 5.4 $\pm$ 0.2 & -\tablenotemark{\tiny l} & --249 & -\tablenotemark{\tiny l}\\
        {He I} & 5876 & 5875.621 & { 1.71$\times$10$^8$} & 186,101.55 & -\tablenotemark{\tiny k} & 1.1 $\pm$ 0.1 & 4.4 $\pm$ 0.5 & --249 & -\tablenotemark{\tiny g}\\ \hline
\enddata

\tablenotetext{a}{We use the numbers in this column to label the individual emission lines throughout the paper.
}
\tablenotetext{b}{
From the NIST Atomic Spectra Database \citep{NIST_ASD}.
}
\tablenotetext{c}{The revised critical density {for an electron temperature of 10$^4$ K}. See Section \ref{sec: discussion calc} for the definition and how we obtained the values.}

\tablenotetext{d} {HVC1 intensities measured by integrating the emission from 1" to 1".5 from the star and averaging over --270 to --230 km s$^{-1}$.
The emission is blended with that from HVC2 for all the forbidden lines except  [S II] 6731 Å, 6716 Å, and [N II] 6583 Å, or blended with that of magnetospheric accretion or a compact inner wind for H$\alpha$ and H$\beta$. See Section \ref{sec: PV}} for details.
\tablenotetext{e}{HVC2 intensities measured by integrating 
the emission within $\pm$ 0".25 of the intensity peak
and averaging it from --205 to --165 km s$^{-1}$ for each line.}
\tablenotetext{f}{Measured by fitting 
the intensity profile over a range of 40 spectral pixels ($\Delta V \sim 40$ km s$^{-1}$) using
a quadratic function.
The intensity profile for HVC1 is extracted from the PV diagram by spatially integrating the emission from 1\arcsec to 1\farcs5 from the star.
That for HVC2 is extracted from the distance from the peak position to where the intensity drops to 30~\% of the peak value at the continuum (Section \ref{sec: major}).
}
\tablenotetext{g}{
A peak is not seen or has a low signal-to-noise.
}
\tablenotetext{h}{We cannot make a reliable measurement due to contaminating emission from the [O II] 7331 Å line as well as the bright HVC2 emission.} 
\tablenotetext{i}{The HVC2 may be blended with another line.} 
\tablenotetext{j}{Difficult to accurately subtract the stellar continuum and/or absorption.}
\tablenotetext{k}{The revised critical density cannot be calculated as ChiantiPy and Pyneb (see Section \ref{sec: discussion calc}) include only thermal collisions when calculating the line intensities. However, the production of H$\alpha$, H$\beta$, and He I also depends on (1) photoionization by UV emission from upstream \citep[see, e.g.,][]{Hollenbach89}; and (2) the recombination of H$^+$ and He$^+$ ions.}
\tablenotetext{l}{The emission may be from magnetospheric accretion or inner winds.}

\end{deluxetable*}
\end{longrotatetable}

\section{Results} \label{sec: results}

We identified 26 emission lines, including 23 forbidden lines and H$\alpha$, H$\beta$, and He I 5876 Å, all associated with the outflowing gas. In Section \ref{sec: PV}, we summarize the properties of the observed emission lines.
In Section \ref{sec: major}, we show the intensity profiles and position spectra of our major target lines (i.e., [O I] 6300, 5577 Å, [S II] 6731 Å, and [N II] 6583 Å) and show the features of their identified LVCs.

In Section \ref{sec: MGF}, we present the Gaussian decomposition of the [O I] 6300 Å line profile and investigate whether this technique allows us to decompose flow components with different physical origins.
We select this line as it is the most thoroughly investigated optical forbidden line associated with jets and disk winds in T-Tauri stars \citep{Eisloffel00_PPIV, Ray07, Frank14}.
The analysis of other lines will be included in a separate paper (Otten et al. in prep).


All the velocities described in this section are radial velocities directly measured from the observations. The use of deprojected velocities is avoided in this section as these can suffer from the measurement uncertainty in the inclination angles of the flow \citep[e.g.,][]{Takami23}.

\subsection{Overview of the emission lines properties} \label{sec: PV}

\begin{figure*}[ht]
  
  \includegraphics[width=1.0\textwidth]{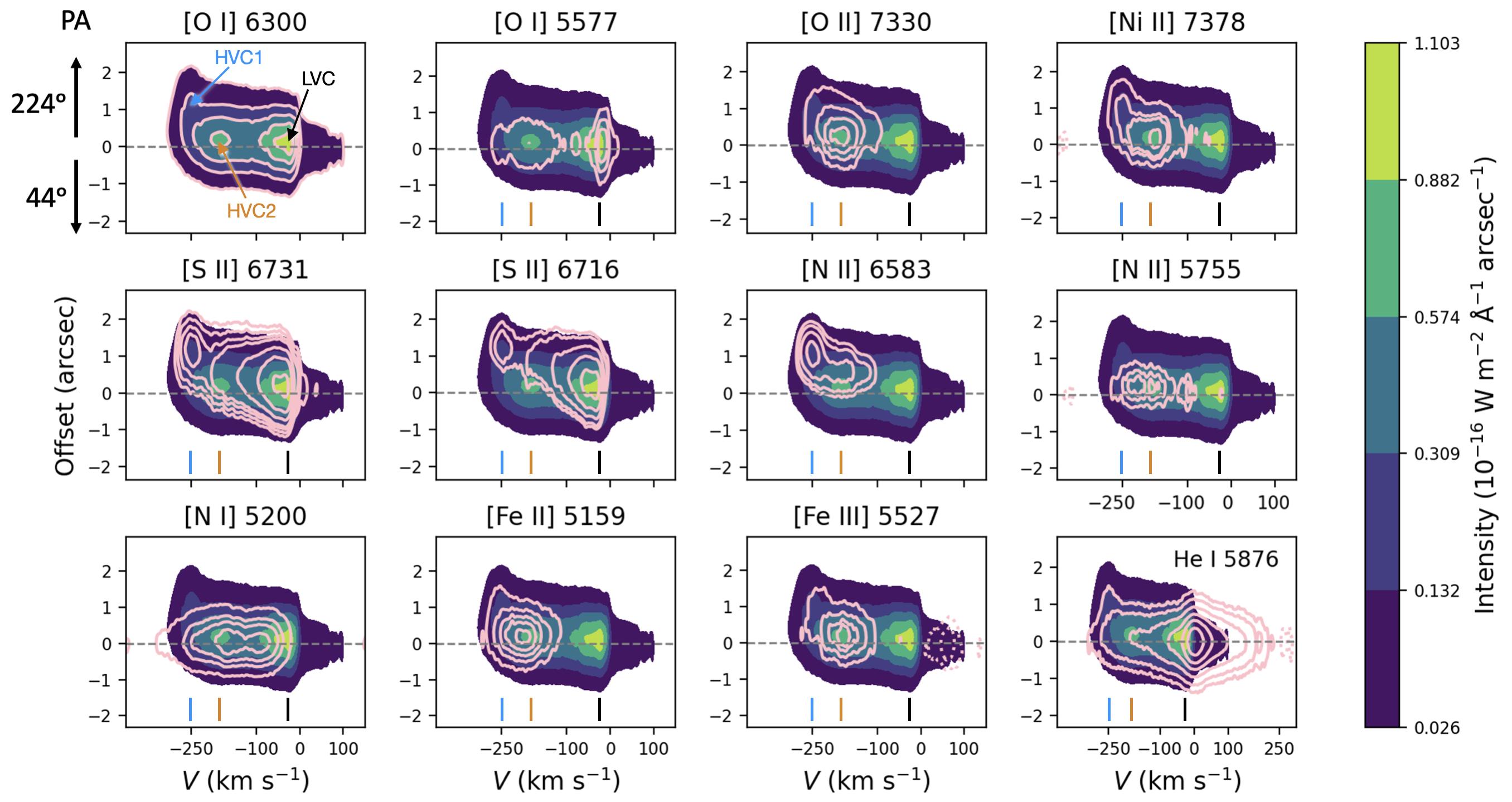}
  \centering
  \caption{
  The PV diagram for 12 emission lines in \textit{pink} contour lines plotted on top of the color contoured [O I] 6300 Å line as a reference.
  The contour levels for each emission line 
  show the presence or absence of the spatially extended HVC (HVC1, $v \sim -$250 km s$^{-1}$), the HVC at the base (HVC2, $v \sim -$185 km s$^{-1}$), and the LVC ($v > -$100 km s$^{-1}$). 
  The blue, orange, and black vertical marks at the bottom of each panel denote the velocities of the HVC1, HVC2, and LVC, respectively, 
  as measured
  for the [O I] 6300 Å line.
  The dotted contours show the residual of the continuum subtraction or perhaps the contamination from the other lines.
  The y-axis shows the spatial extent for PA 224\degree---44\degree.
  The zero point on the y-axis indicates the position of the central star (the gray horizontal dashed lines).}
  \label{PVDfinal}
\end{figure*}

Figure \ref{PVDfinal} shows the PV diagrams for 12 emission lines.
We applied 2-D Gaussian convolution to each PV diagram to increase the signal-to-noise ratio. After the convolution, the angular and velocity resolutions are 1\farcs1 and 6.7 km s$^{-1}$, respectively.
The PV diagrams for the different emission lines are shown in pink contour lines and, for comparison, are overplotted with the colored contour PV diagram of the [OI] 6300 Å line.

%
The remaining 14 lines are not included in the figure for the reasons given below. First, we exclude the [O I] 6364 Å line as it is identical to the [O I] 6300 Å line (but with different intensity levels) due to the same upper energy level. Secondly, H$\alpha$ and H$\beta$ are dominated by bright emission at the star due to magnetospheric accretion or an inner wind very close to the star \citep{Najita00,Hartmann16}. The remaining lines are either blended with other permitted lines, or are too faint compared to the residual of the stellar absorption or continuum emission.\\

We identified three components in the forbidden line emissions of Figure \ref{PVDfinal}.
First, many lines exhibit a spatially extended component at $v \sim -$250 km s$^{-1}$ (hereafter HVC1). Some lines also exhibit two more components near the stellar position centered at a high velocity ($v \sim -$185 km s$^{-1}$; hereafter HVC2) and low velocity ($v > -$100 km s$^{-1}$; hereafter LVC). 
As discussed in later sections, HVC1 and HVC2 are associated with a collimated jet.
Below, we explain the individual components shown in Figure \ref{PVDfinal} in detail.\\


HVC1 is observed as a distinct emission component in the [S II] 6716, and 6731 Å lines and the [N II] 6583 Å line, peaking at about 1\farcs2 from the star.
HVC2 is observed as a distinct component in all the forbidden emission lines in Figure \ref{PVDfinal} except for the above [S II] and [N II] lines. Some of the lines with HVC2 show extended emission at the velocity of HVC1 but without a clear peak as in these [S II] and [N II] lines, probably because the emission is blended with the significantly brighter HVC2. In contrast, the [S II] 6716, 6731 Å lines and the [N II] 6583 Å line do not exhibit a clear peak for HVC2, as described below in detail.
In Table \ref{tab: line} (from the second to last column), we added the measured intensity
 by integrating the emission from 1\arcsec to 1\farcs5 from the star and averaging over $-$270 to $-$230 km s$^{-1}$.
In total, Table \ref{tab: line} lists 26 emission lines we have identified.
In addition, a line near 5334 Å and a line near 5376 Å were detected, but we could not identify the corresponding ions and transitions.\\

All of the forbidden emission lines primarily show
blueshifted components, as these are associated with a jet and winds, and the counter jet and/or winds are obscured by an optically thick circumstellar disk \citep[e.g.,][for a review]{Eisloffel00_PPIV}.
In contrast, the He I 5876 Å emission line in Figure \ref{PVDfinal} shows a prominent redshifted emission and a bright emission peak near zero velocity, most likely associated with magnetospheric mass accretion in the inner disk \citep{Najita00}.
This is probably because, as for optical permitted line emission in many cases, the emission is primarily associated with magnetospheric mass accretion \citep[][for reviews]{Najita00, Hartmann16}. 
Nevertheless,
the PV diagram for the He I 5876 Å line still shows extended emission at the velocity of HVC1 ($v \sim -250$ km s$^{-1}$), indicating the presence of this emission component.
Furthermore, the emission at the velocity of HVC2 ($v \sim -185$ km s$^{-1}$) is brighter than that at $v \sim +185$ km s$^{-1}$, indicating the presence of HVC2 emission as well.
\\

\begin{figure}[ht]
  \centering
  \includegraphics[width=8cm]{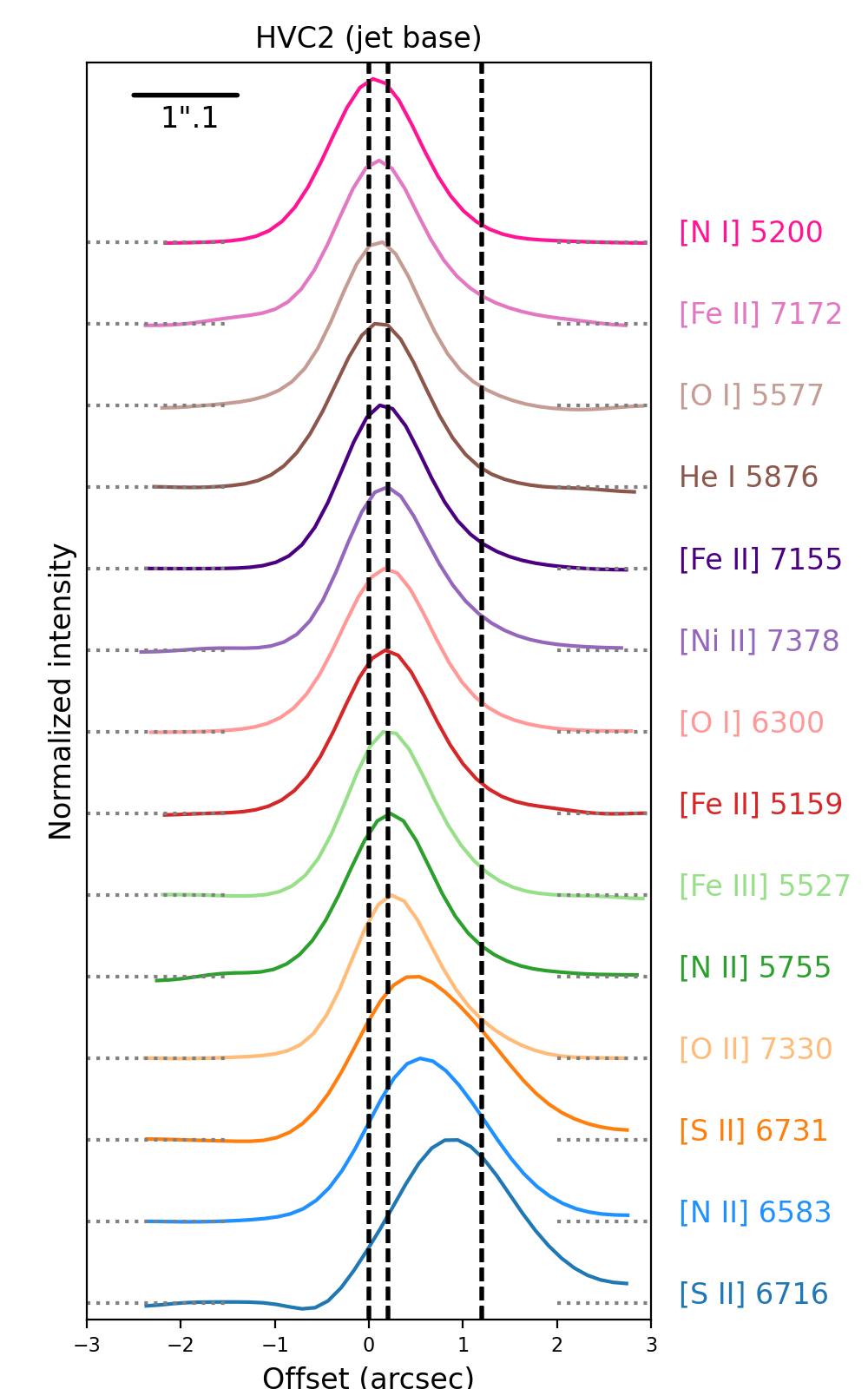}
  \caption{The spatial profiles of HVC2.
  The profiles are made by averaging the spatial cuts of the PV diagrams from $v = -$205 to $-$165 km s$^{-1}$, and the peak intensities are normalized to unity. 
  The profiles are arranged from top to bottom with increasing offset of the peak intensities.
  The black bar in the upper left corner shows the angular resolution of the profiles, which is 1".1.
  The vertical black dashed lines denote the offsets at 0", 0".2, and 1".2 from the central star towards the jet direction.
  The horizontal gray dotted lines denote the zero intensities for each emission line.}
  \label{HVC2_spatial}
\end{figure}

Figure \ref{HVC2_spatial} shows the observed spatial profiles for HVC2
and the relative offset with respect to the central star position.
The profiles are extracted by averaging the spatial cuts of the PV diagrams from $v =$ --205 to --165 km s$^{-1}$, with the peak intensities normalized to unity. 
We include all the emission lines with reliable measurements, i.e., without contamination from the other emission lines or the residual of the subtraction of the stellar absorption lines.
The third last column in Table \ref{tab: line} shows the HVC2 intensities measured by integrating the emission within $\pm$ 0".25 of the intensity peak and averaging from --205 to --165 km s$^{-1}$ for each line.

Except for the [S II] lines (6731 and 6716 Å) and the [N II] 6583 Å line, all the lines presented in Figure \ref{HVC2_spatial} are spatially unresolved, with a relatively small offset at the intensity peak (0".06 - 0".25).
The [N I] 5200 Å line has the smallest offset, while the [S II] 6716 Å line has the largest offset from the central star towards the jet direction.
The [S II] 6716, 6731 Å lines and the [N II] 6583 Å line are marginally spatially resolved with peak positions of  0".88, 0".48, and 0".57, respectively.
The different spatial extent compared with the other lines
might hint at
a different physical mechanism that produces these emission lines
(see Section \ref{sec: discussion HVC}).\\
%

In the rest of the paper, we will primarily use [O I], [S II], [N II], He I and [Fe II] 5159 Å lines for detailed analysis and discussion. 
In the future, the analysis of other lines from DG Tau and other T-Tauri stars would be useful for investigating detailed shock conditions (see Section \ref{sec: discussion HVC} for more discussion) and quantitatively measuring iron depletion (see Section \ref{sec: discussion LVC}).

\subsection{Major target lines} \label{sec: major}
\begin{figure*}[ht]
  \centering
  \includegraphics[width=1.0\textwidth]{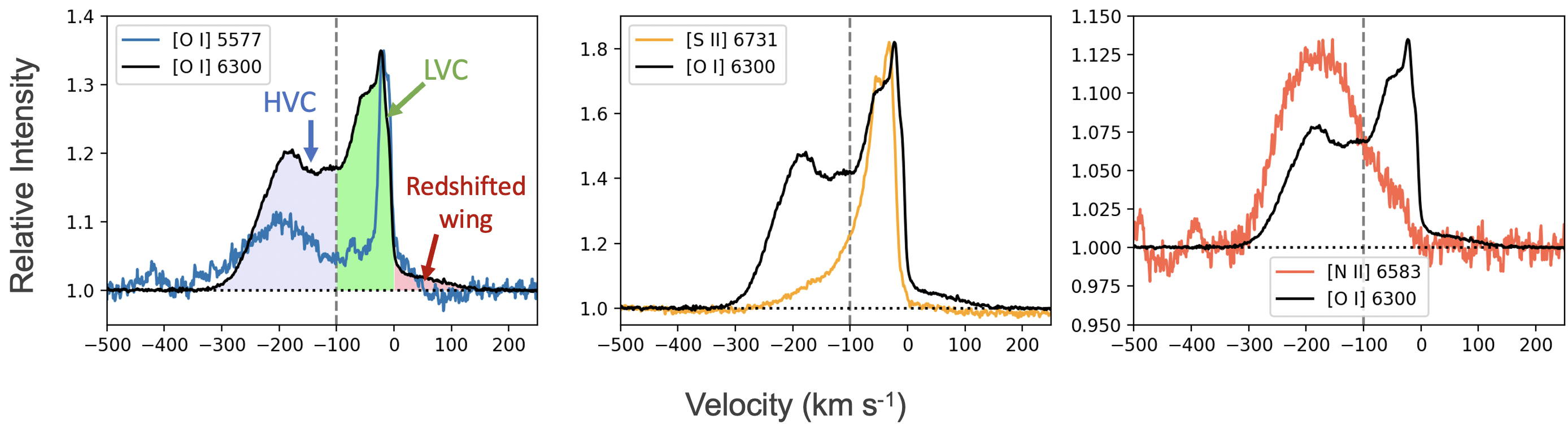}
  \caption{The [O I] 5577 Å, [S II] 6731 Å, and [N II] 6583 Å line profiles 
  superimposed
  by the [O I] 6300 Å line profile. The [O I] 5577 Å, [S II] 6731 Å, and [N II] 6583 Å line profiles are normalized to the continuum. The [O I] 6300 Å line profiles are scaled down to match the peaks of the other lines for comparison. The black dotted lines mark the continuum, and the vertical gray dashed lines denote the boundary between the HVC and LVC we define (see Section \ref{sec: major}). The blue, green, and red shaded parts in the left panel show the HVC, LVC, and redshifted wing, respectively.}
  \label{line profile}
\end{figure*}

\begin{figure}[ht]
  \centering
  \includegraphics[width=8cm]{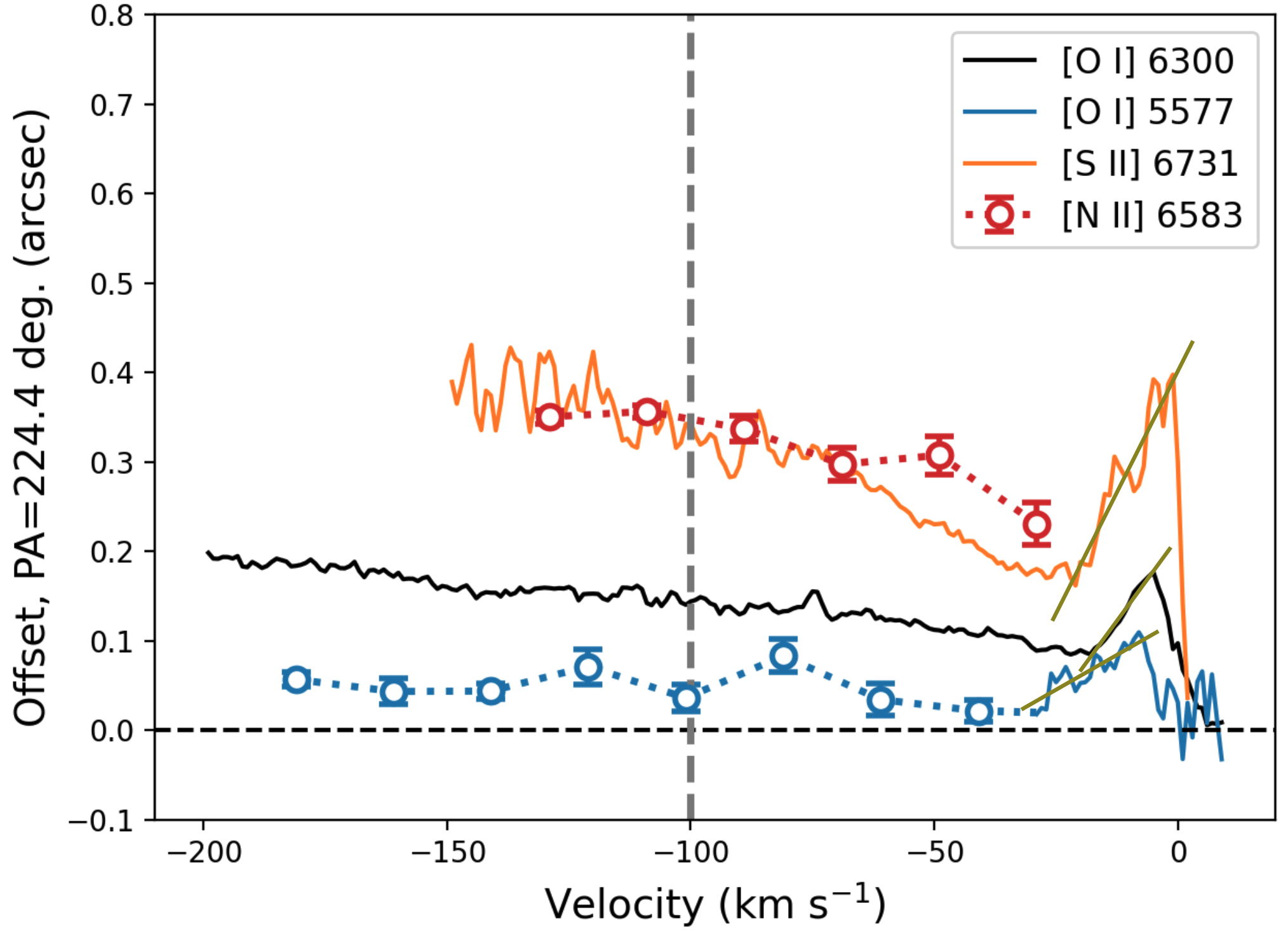}
  \caption{The offset position spectra of the  [O I] 6300 Å, [O I] 5577 Å, [S II] 6731 Å, and [N II] 6583 Å lines with respect to the PA of the jet \citep{Kepner93,Lavalley97,Takami23}.
  The dotted lines with open circles show the position spectra binned every 20 km s$^{-1}$ to increase the signal-to-noise ratio
  for the [O I] 5577 Å and [N II] 6583 Å lines.
  The green solid lines
  at $\sim$--30 to $\sim$--5 km s$^{-1}$
  indicate the presence of the negative velocity gradients discussed in Sections 3.2 and 4.3.
  The black dashed horizontal line marks the stellar position,
  and the gray dashed vertical line denotes the boundary between the HVC and the LVC.}
  \label{position}
\end{figure}

\begin{figure}[ht]
  \centering
  \includegraphics[width=8cm]{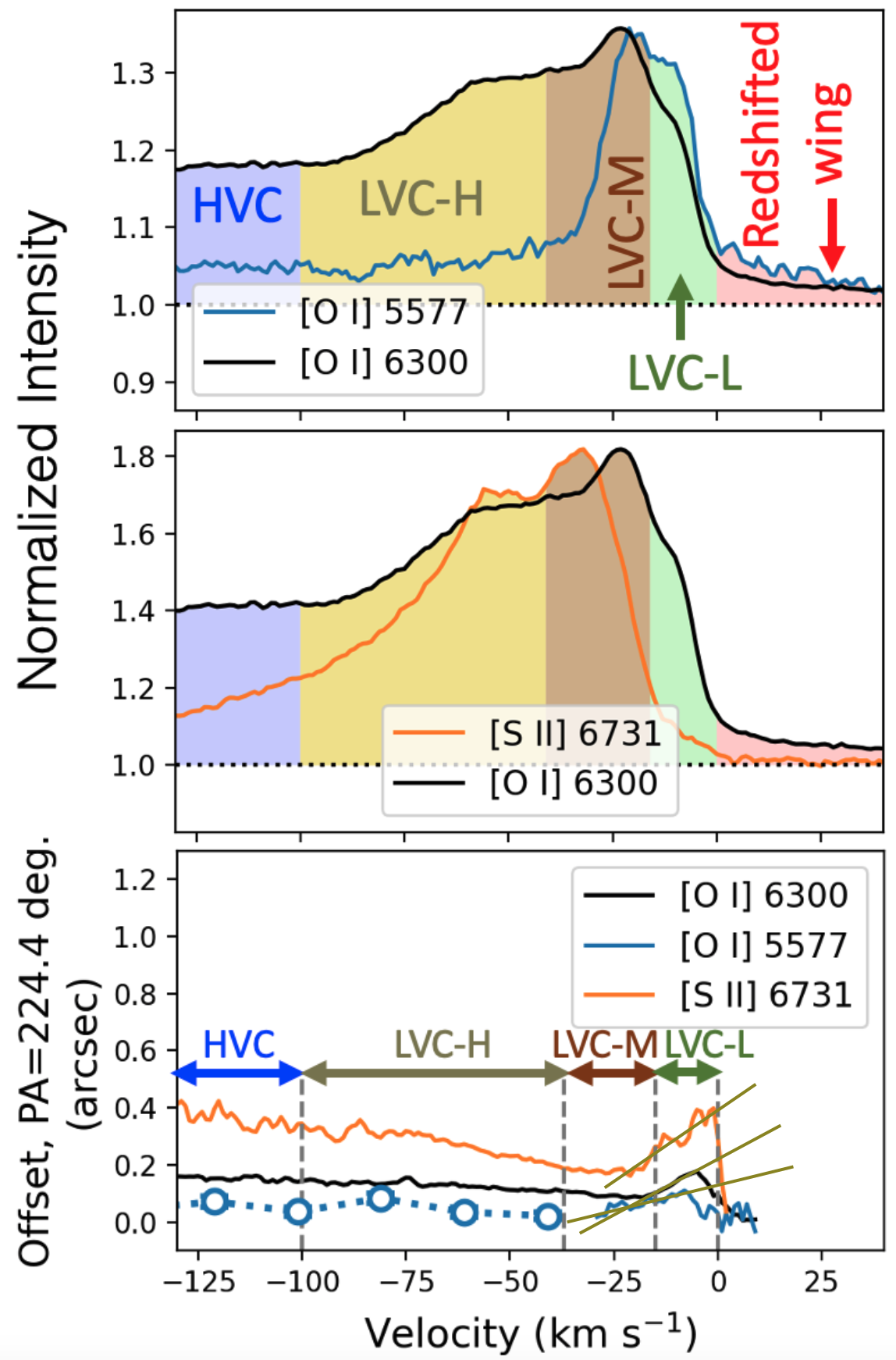}
  \caption{The same line profiles and the position spectra of the [O I] 6300 Å, [O I] 5577 Å, [S II] 6731 Å lines presented in Figure \ref{line profile} and \ref{position}, but focused on the LVC.}
  \label{3in1LVC}
\end{figure}

\begin{figure}[ht]
  \centering
  \includegraphics[width=8.5cm]{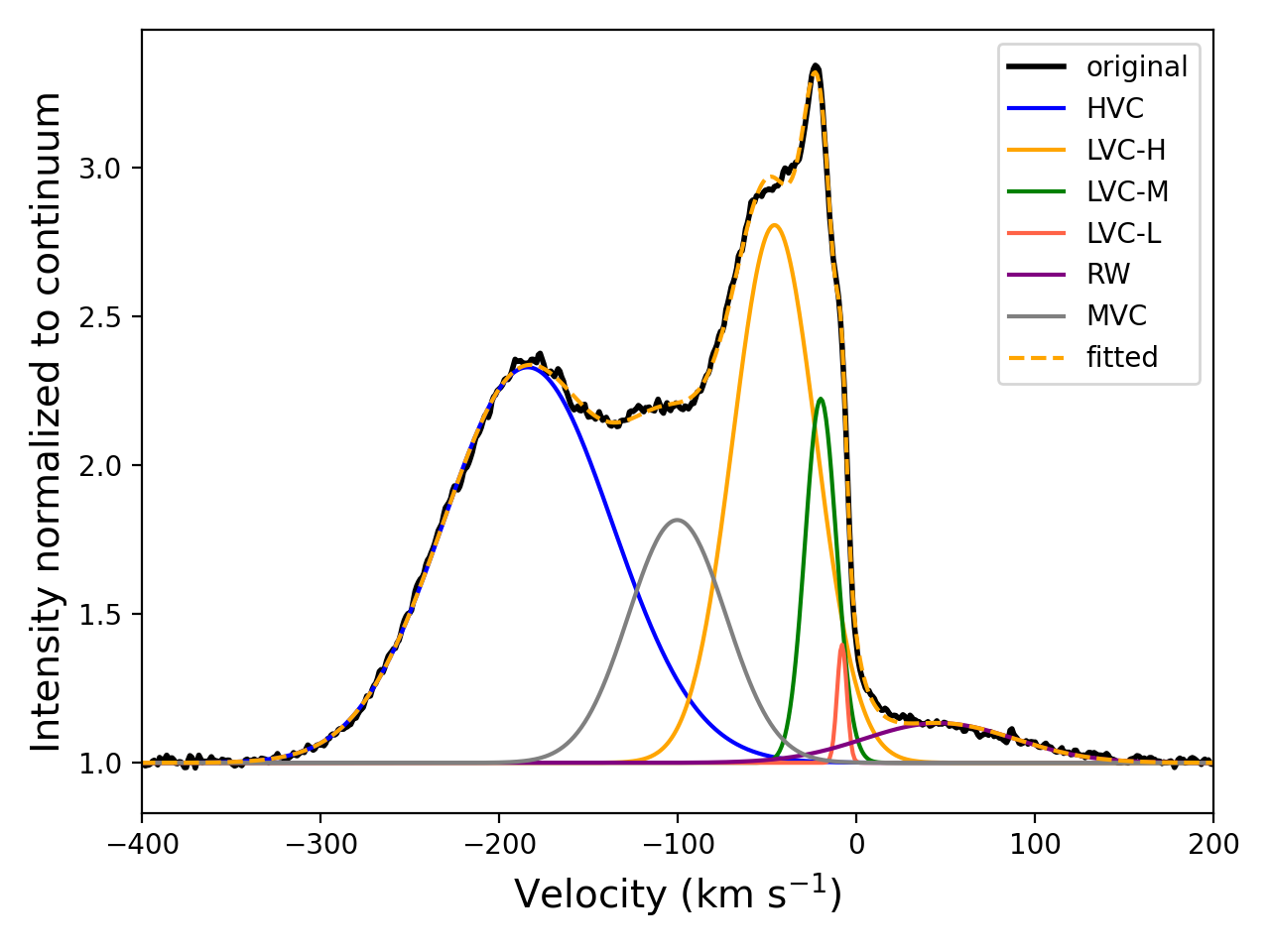}
  \caption{The multi-Gaussian fit result of the [O I] 6300 Å line profile. It is decomposed into six components: HVC, LVC-H, LVC-M, LVC-L, the redshifted wing (``RW''), and MVC.}
  \label{MGF6300}
\end{figure}

\begin{figure}[ht]
  \centering
  \includegraphics[width=8.5cm]{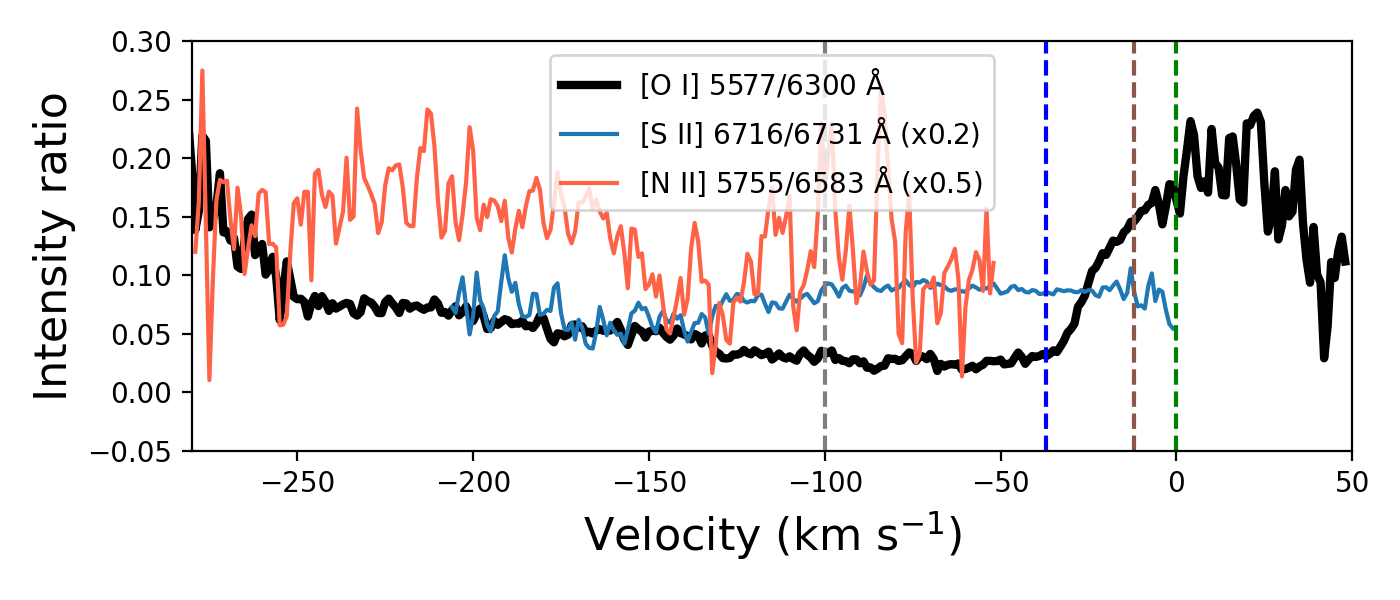}
  \caption{The intensity ratio profiles of [O I] 5577/6300, [S II] 6716/6731, and [N II] 5755/6583 corrected for extinction ($A_V = 0.8$). The gray, blue, brown, and green dashed lines mark the lower boundaries of the HVC, LVCH, LVC-M, and LVC-L, respectively.}
  \label{LR profiles}
\end{figure}

\begin{deluxetable}{lcc}
\tablewidth{0pt} 
\tablecaption{
Individual
Velocity
Components
in This Study.
\label{tab:velocity range}}
\tablehead{
\colhead{Component} & \colhead{Velocity range\tablenotemark{\tiny a} (km s$^{-1}$)} 
} 
\startdata 
HVC\tablenotemark{\tiny  b} & $v < -100$\\
LVC-H & $-100 < v < -41$\\
LVC-M & $-41 < v < -16$\\
LVC-L & $-16 < v < 0$\\
Red-shifted wing & $v > 0$\\
\enddata
\tablenotetext{a}{Velocities at which we identify the peaks and humps in the line profiles in Figures \ref{line profile} and \ref{3in1LVC}.}
\tablenotetext{b}{Dominated by HVC2 in the PV diagrams in Figure \ref{PVDfinal} (Section \ref{sec: PV}) due to the spatial range over which we extracted the line profiles (see Section \ref{sec: major}).}
\end{deluxetable}

Figure \ref{line profile} shows the intensity profiles for the
[O I] 6300 and 5577 Å, [S II] 6731 Å, and [N II] 6583 Å lines, i.e., the major lines due to the high signal-to-noise detections compared to the rest of the lines.
The intensity profiles are extracted from the 2-D spectra by integrating from the peak position to where the intensity drops to 30~\% of the peak value at the continuum.
The line profiles show HVC and LVC at $|v| \gtrsim 100$ km s$^{-1}$ and $\lesssim 100$ km s$^{-1}$, respectively.
The [O I] 6300 Å line profile shows a redshifted wing in addition to these two blueshifted components.

Figure \ref{line profile} \textit{left}
shows that both of the [O I] 6300 Å and 5577 Å lines have HVC (dominated by HVC2 in Figure \ref{PVDfinal}), LVC, and redshifted wings, but with different line shapes. 
{The [S II] 6731 Å emission in Figure \ref{line profile} \textit{middle} shows a prominent LVC,}
and its line profile shape is similar to that of the [O I] 6300 Å line. However,
the [S II] 6731 Å line profile in this panel does not show
a clear peak corresponding to HVC1 or HVC2.
This is because these emission components
are located outside of the spatial range where we extracted the line profile (Figure \ref{PVDfinal}). Furthermore, 
the [S II] 6731 Å line profile in this panel does not show a redshifted wing like the [O I] 6300 Å and 5577 Å line profiles shown in the left panel of the same figure.
{The [N II] 6583 Å line in Figure \ref{line profile} \textit{right}}
is dominated by the HVC, and it may have a marginal detection of the LVC. Note that we scale the [O I] 6300 Å line profile to match the maximum intensities of the other lines for comparison.\\

Figure \ref{position} shows the offset position spectra of the [O I] 6300, 5577 Å, [S II] 6731 Å, and [N II] 6583 Å lines made using the SA method (Section \ref{sec: method}). 
For the velocity ranges shown by the open circles and dotted lines, we binned the spectra with a bin of 20 km s$^{-1}$ to increase the signal-to-noise ratios.
The velocity ranges that are not plotted for each offset position spectrum are due to the following reasons:
(1) the measurements at high velocities are not reliable due to 
either offsets that are too large, or a spatial profile that substantially deviates from a Gaussian profile, and (2) the offset position spectrum at low velocities suffers from high noise due to small line-to-continuum ratios (see Section \ref{sec: method} for details).\\

In Figure \ref{position}, the [N II] 6583 Å and [S II] 6731 Å lines show positional offsets larger than those of the [O I] lines.
The [O I] 5577 Å line has the smallest offset (0\farcs06 - 0\farcs1),
so it may trace the base of the jet and winds. 
At $v < -$30 km s$^{-1}$, the positional offset for the [S II] and [O I] 6300 Å lines increases with increasing radial flow velocity |$v$|. 
The positional offset of the [O I] 5577 Å line does not show any clear correlation with its radial velocity.
At velocities of $\sim$--30 to $\sim$--5 km s$^{-1}$, the positional offset decreases as the radial
blueshifted
velocity |$v$| increases.
This trend is referred to as ``the negative velocity gradient'' first defined by \citet{Whelan21} in the offset position in the spectra for RU Lupi using the SA technique.

In order to see more in detail the LVC region in the spectra,
Figure \ref{3in1LVC} shows the line profiles and the position spectra of the [O I] 6300 Å, [O I] 5577 Å, and [S II] 6731 Å lines, focusing on their LVCs.
The [O I] 6300 Å line profile peaks at $v$ = --23 km s$^{-1}$, with two humps at $v$ = --100 to --41 and --16 to 0 km s$^{-1}$, respectively, indicating the presence of three distinct emission components. For the rest of the paper, we label these 
LVC components as
LVC-H, LVC-M, and LVC-L for high to low flow velocities.
Table \ref{tab:velocity range} summarizes the velocity ranges over which we \textit{identified} the individual emission components.
In practice, it is highly unlikely that the presence of each emission component is limited to the tabulated velocity range. Each of the tabulated emission components must extend beyond the tabulated velocity range and blend with the adjacent emission components (Section \ref{sec: MGF}).

In the top panel of Figure \ref{3in1LVC},
the [O I] 5577 Å emission clearly shows the presence of LVC-M and LVC-L. 
However, the profile of this emission line does not show the presence of a hump at the velocity range of LVC-H (--100 to --41 km s$^{-1}$, see Table \ref{tab:velocity range}), suggesting the absence of this velocity component. For this line, the emission observed at this velocity range would instead be the wings of HVC and LVC-M extending over the velocity range tabulated in Table \ref{tab:velocity range}.
In the middle panel of Figure \ref{3in1LVC},
the [S II] 6731 Å emission shows clear LVC-H and LVC-M, but 
its line profile
does not
show the presence of
excess emission corresponding to LVC-L.
The [S II] 6731 Å LVC-M velocity peaks at $-$32 km s$^{-1}$, which is slightly more blueshifted than that of the [O I] 6300 Å LVC-M at $-$23 km s$^{-1}$.\\

The bottom panel of Figure \ref{3in1LVC} shows that the negative velocity gradient described above lies in 
the velocity ranges
for
LVC-L and LVC-M
tabulated in Table \ref{tab:velocity range}.
This gradient is probably due to the emission associated with LVC-M
for the reasons below. The [S II] 6731 Å line, which clearly shows
the negative velocity gradient in its offset position spectrum, does not clearly show the presence of the LVC-L emission in its line profile
as described above. Furthermore, the LVC-L emission is marginal in the line profile of [O I] 6300 Å, which also shows the negative velocity gradient
in its offset position spectrum. 
Therefore, the LVC-M emission contributes a larger velocity range within the negative velocity gradient in the position spectrum.

Additionally, we rule out the possibility that the negative velocity gradient is due to one of the other observed velocity components (LVC-H, HVC2, and HVC1 from low to high flow velocities) for the following reasons.
This velocity gradient is observed in the [O I] 5577 Å and [S II] 6731 Å lines, which do not clearly show the presence of LVC-H and HVC2, respectively. The velocity for HVC1 ($v \sim -$250 km s$^{-1}$), which is associated with the extended jet (Sections \ref{sec: PV} and \ref{sec: discussion HVC}) is very different from that of the gradient ($v \sim -$20 to 0 km s$^{-1}$).

\subsection{Gaussian decomposition of the emission components}\label{sec: MGF}

Previous studies attempted to decompose the HVC and two components in the LVC (BC/NC, see Section \ref{sec:intro}) using three Gaussians \citep{Rigliaco13, Natta14, Simon16, Banzatti19}.
Similarly, we take each line profile as a composite of Gaussian functions. We developed a Python program to carry out the multi-Gaussian fit. After inputting the initial guesses of centroid velocities, heights, and FWHMs for each possible Gaussian component, 
{\fontfamily{qcr}\selectfont scipy.optimize.leastsq} will calculate the optimized parameters. Following \citet{Banzatti19}, we determined the number of Gaussian components to describe a line profile by calculating improvement in the reduced chi-square. We continued adding components until the reduced chi-square was not improved by 20~\% or until there was a component with a similar peak height to the continuum noise.\\

Figure \ref{MGF6300} shows the best multi-Gaussian fitting result for the [O I] 6300 Å line profile.
It is clear that the HVC, LVC-H, LVC-M, LVC-L, and the redshifted wing have their corresponding Gaussian components after the decomposition.
As explained in Section 3.2, these three LVC components are highly blended.
In addition, we found an MVC at --100 km s$^{-1}$, which is more blueshifted than reported by \citet{Giannini19}.
The centroid velocities, heights, and FWHMs for each Gaussian component are presented in Table \ref{tab: MGF6300}.\\

Figure \ref{LR profiles} shows the intensity ratios for [O I] 5577/6300, [S II] 6716/6731, and [N II] 5755/6583 as a function of velocity. 
The intensity ratio profiles are corrected for an extinction of $A_V =$ 0.8 $\pm$ 0.8 (see Section \ref{sec: discussion LVC} for details).
Such an extinction correction increases the [O I] 5577/6300 and [N II] 5755/6583 intensity ratios by about 10\%. 
Still, the overall trends before and after the correction are the same.
The figure does not show clear evidence for a variety of gas components with different physical conditions corresponding to the Gaussians shown in Figure \ref{MGF6300}. Instead, the [O I] 5577/6300 ratio shows an excess at $\gtrsim$ --40 km s$^{-1}$, indicative of the presence of a single emission component with a high density or temperature (see Table \ref{tab: line}) at this velocity range.

\begin{deluxetable}{llll}
\tablewidth{0pt} 
\tablecaption{Best fit parameters of each velocity component 
for the [O I] 6300 Å line profile.
\label{tab: MGF6300}}
\tablehead{
\colhead{Component} & \colhead{$v_{\mathrm{c}}$ (km s$^{-1}$)\tablenotemark{\tiny a}} & \colhead{$I$\tablenotemark{\tiny b}} & \colhead{FWHM (km s$^{-1}$)}  
} 
\startdata 
HVC & $-184$ & $1.3$ & $111$\\
LVC-H & $-46$ & $1.8$ & $55$\\
LVC-M & $-20$ & $1.2$ & $21$\\
LVC-L & $-8$ & $0.4$ & $6$\\
RW & $46$ & $0.1$ & $99$\\
MVC & $-100$ & $0.8$ & $64$\\
\enddata
\tablenotetext{a}{Centroid velocity}
\tablenotetext{b}{Intensity peak for the Gaussian relative to the continuum} 
\end{deluxetable}

\section{Physical Conditions and the Nature of Individual Emission Components} \label{sec: discus}

In Section \ref{sec: discussion calc}, we explain how we use ChiantiPy and Pyneb to investigate the physical conditions (electron density and its spatial distribution, electron temperature, and inferred mass loss rates) of the outflow. In Section \ref{sec: discussion HVC}, we discuss the origins of the HVCs. In Section \ref{sec: discussion LVC}, we discuss the possible origins of the individual LVC sub-components.
In Section \ref{sec: MLR}, we estimate and discuss the wind mass-loss rate.

\subsection{Calculations using ChiantiPy and Pyneb}\label{sec: discussion calc}

To investigate the physical conditions in the jet and winds in detail, we use ChiantiPy\footnote{https://chiantipy.readthedocs.io/en/latest/}, a Python package, to calculate astrophysical spectra using the CHIANTI atomic database \citep{DelZanna21}.
This software allows us to calculate the intensities of the transitions of interest for any given density and temperature with thermal processes, including excitation and de-excitation for the transitions, and ionization and recombination for the corresponding atoms and ions. 
For [Fe II], CHIANTI does not include all the energy levels; instead, we used Pyneb\footnote{https://pypi.org/project/PyNeb/} \citep{Luridiana15} to obtain the calculations for thermal excitation and de-excitation.
The versions of the software and the database we use are 0.15.0, 10.0.2, and 1.1.18 for ChiantiPy, CHIANTI, and Pyneb, respectively.\\

The critical density, a well-known emission line property, is conventionally defined as the density where spontaneous emission is balanced by collisional de-excitation. The critical density has been extensively used to concisely discuss electron densities for emission line regions, as the emission is enhanced above the critical density at which the gas reaches local thermal equilibrium (LTE). In this context, transitions with higher critical densities tend to be associated with regions of higher electron densities \citep[cf.][]{Stahler04}. However, the dependency of the line intensity on the electron density is more complicated 
for some cases. For example, each energy level for [Fe II] has two critical densities due to two different groups of transitions: i.e., (1) those associated with different fine spin-orbit split levels for the total angular momenta; and (2) those associated with more separated terms for differing orbital and spin angular momenta.
Hence, we defined a ``revised critical density'' as follows \citep[cf.][]{Takami10b}:
\begin{equation}
{I(n_{\mathrm{crit}}{, T_e})} \equiv 0.5 \times 
I_\mathrm{LTE}{(T_e)}, \label{eq:n_crit_rev}
\end{equation}
where $I_\mathrm{LTE}$ is the intensity at the local thermal equilibrium (LTE){; and $T_e$ is the electron temperature}.
For our calculations, we use a very high electron density (10$^{12}$ cm$^{-3}$) 
 such that all the energy levels related to the transition of interest reached LTE.

{
In Table \ref{tab: line}, we tabulate the revised critical densities for $T_e$=10$^4$ K, a typical temperature of the forbidden line emission for the jet/wind from T-Tauri stars \citep[e.g.,][]{Dougados02,Fang18,Weber20,Murphy24}. In Appendix \ref{app:n_crit}, we perform comparisons between the conventional and revised critical densities, and discuss the validity of the latter.
}

\subsection{The origin of the HVCs}\label{sec: discussion HVC}

Figure \ref{PVDfinal} shows that both HVC1 and HVC2 are associated with the [O I] and He I lines. This fact implies that the gas in these regions contains O$^0$ and He$^+$, considering that the He I line(s) originates via recombination. These two lines require gas with remarkably different excitation conditions, as explained below. The presence of He$^+$ requires the ionization of He$^0$ with an ionization energy of 24.6 eV. However, the electron or a photon required for such ionization would easily ionize O$^0$, whose ionization potential is 13.6 eV. The coexistence of such gases can be explained with shocks, as discussed in previous studies \citep{Lavalley00, Bacciotti00, Takami02b}. In shocks, the He I line is generated via gas cooling in the postshock region, where the He$^{+}$ is recombined in the cooling process, and the O I atoms are the major atomic coolant \citep{Bacciotti99, Hollenbach89}.
The large range of gas temperatures associated with these emission components is also corroborated by the detection of low-to-high excitation lines such as [N I], [N II], [O III], [S II], [Ca II], [Fe II], [Fe III], [Ni II], H$\alpha$, and H$\beta$ lines \citep[Section \ref{sec: PV}; see also][for their excitations in shocks]{Hollenbach89,Hartigan00_PPIV}.

We interpret HVC1 as a moving knot of DG Tau A's collimated jet based on its large offset from the central star toward the jet direction (see Figure \ref{PVDfinal}). The existence of such ``moving knots'' has already been reported by previous studies \citep{Pyo03, White14a, Takami23}.
The HVC2 peaks are at positions offset by
0".06 - 0".25 from the central star toward the jet (see Section \ref{sec: PV}). This component may be associated with the stationary shock discussed in several previous studies \citep{Schneider08, Schneider11, Gudel11, White14b, Takami23}.
This is an emission component close to the central star located on the jet axis, which appears to be stationary over time.
The observations by \citet{Gudel11} and \citet{White14b} showed that the stationary knot is located at about 0".2 from DG Tau A, which is consistent with our result.
According to the model proposed by \citet{Gunther14}, 
this stationary shock component can be caused by the recollimation of the fast stellar wind by the magnetic and thermodynamic pressure from the 
low-velocity
disk winds.

\begin{figure*}[ht]
  \centering
  \includegraphics[width=18cm]{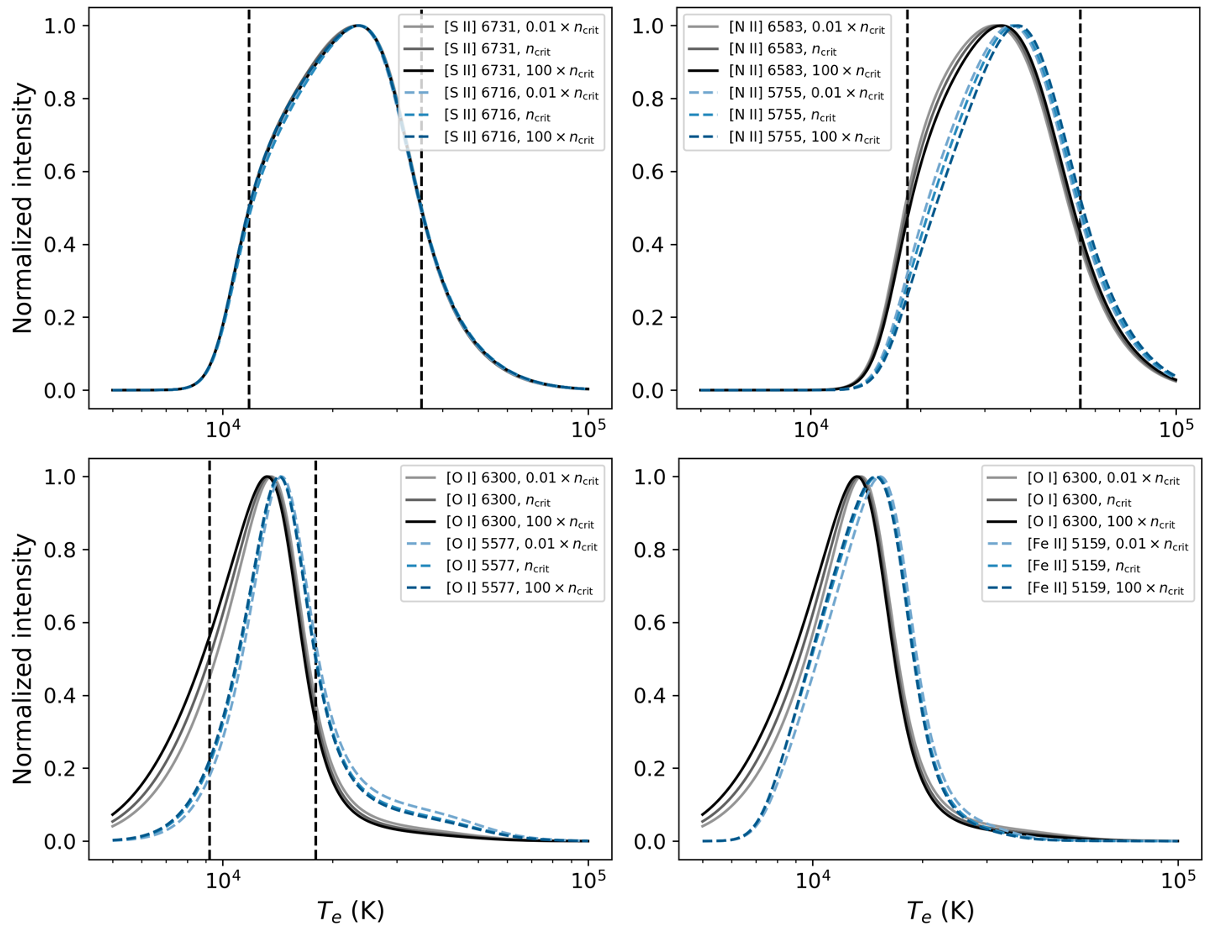}
  \caption{The calculated intensity as a function of temperature at three different electron densities (0.01$\times$$n_\mathrm{crit}$, $n_\mathrm{crit}$, and 100$\times$$n_\mathrm{crit}$, using values of $n_\mathrm{crit}$ listed in Table \ref{tab: line}) for [S II] 6731, 6716 Å (upper left), [N II] 6583, 5755 Å (upper right), [O I] 6300, 5577 Å (lower left), and [O I] 6300 Å, [Fe II] 5159 Å (lower right). Each curve is normalized to the peak intensity. 
  In the top-left, top-right, and bottom-left panels,
  the vertical dashed lines denote the temperatures approximately corresponding to half the maximum intensity
  (the ``quasi-characteristic temperatures"; see Section \ref{sec: discussion HVC}).
  }
  \label{IvsT_comp}
\end{figure*}

\begin{figure*}[ht]
  \centering
  \includegraphics[width=18cm]{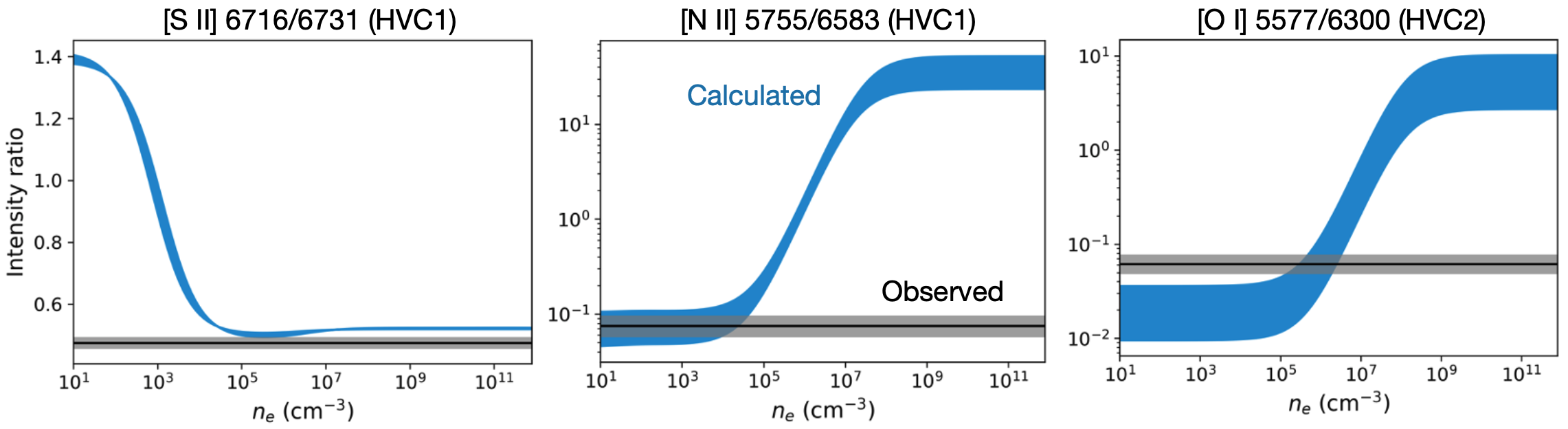}
  \caption{
  The [SII] 6716/6731, [NII] 5755/6583, and [OI] 5577/6300 intensity ratios as a function of the electron density calculated by ChiantiPy (the blue shaded regions). 
  The horizontal black lines represent
  the measured intensity ratios of HVC1 (the left and middle panels) and HVC2 (the right panel) corrected for an extinction of $A_V = 0.8$. 
  The gray bars
  show the uncertainties of the intensity ratio measurements, including the uncertainties in extinction with $A_V = 0.8 \pm 0.8$.
  \label{find hvc density}}
\end{figure*}

We estimate the electron densities for HVC1 and HVC2 using ChiantiPy as follows.
All four panels in Figure \ref{IvsT_comp} show the calculated intensity curves as a function of the electron temperature for three densities: 0.01$\times$$n_\mathrm{crit}$, $n_\mathrm{crit}$, and 100$\times$$n_\mathrm{crit}$. The $n_\mathrm{crit}$ used are those listed in Table \ref{tab: line}.  
Each panel in Figure \ref{IvsT_comp} shows that
the
line intensity 
(and therefore the emissivity of gas)
is
high
over a specific temperature range.
This range is nearly independent of the electron density (see Appendix \ref{app:f8} for detailed discussion).
In each panel, the intensity curves of the two emission lines are similar,
implying that these lines have a similar temperature dependence.\\

For
the [S II], [N II], and [O I] lines
in Figure \ref{IvsT_comp},
we defined the ``characteristic temperature'' 
($T_\mathrm{c}$)
and ``quasi-characteristic temperatures'' 
($T_\mathrm{q}$)
as the temperatures where the maximum and half-maximum intensity occur, respectively. 
These are defined using the curves for the critical density in each panel.
The quasi-characteristic temperatures
$T_\mathrm{q}$
are denoted by vertical dashed lines in Figure \ref{IvsT_comp}.
As explained below, we will use these parameters to demonstrate that the [S II] 6716/6731, [N II] 5755/6583, and [O I] 5577/6300 intensity ratios are primarily dependent on the electron density for shocked slabs in HVC1 and HVC2. Based on this property, we will estimate the electron densities for these emission components.

In the top-left panel of Figure \ref{IvsT_comp},
the intensity curves for the [S II] 6731 Å and 6716 Å lines are nearly identical, and we used the former to determine the lower and upper $T_\mathrm{q}$. For each of the top-right and bottom-left panels, we used the 
[N II] 6583 Å or the [O I] 6300 Å lines (shown in dark gray) to determine the lower $T_\mathrm{q}$, and the [N II] 5755 Å or the [O I] 5577 Å lines (shown in blue) to determine the upper $T_\mathrm{q}$.
As a result, the temperature range determined by the lower and upper $T_\mathrm{q}$ approximately brackets each curve where the value exceeds 0.5 in the top-left, top-right and bottom-left panels in Figure \ref{IvsT_comp}.

Therefore, if gas with a range of temperatures exists over the line of sight of the observations, such as shocked slabs, the observed line intensity is approximately dominated by the gas over the temperature range shown by the two $T_\mathrm{q}$. 
As a result,
the [S II] 6716/6731, [N II] 5755/6583, and [O I] 5577/6300 intensity ratios will be primarily dependent on the electron density.
For each line ratio, we use the temperature range determined by the lower and the upper $T_\mathrm{q}$ to
visualize
possible systematic errors 
in order to investigate
the electron densities,
as demonstrated below.

Figure \ref{find hvc density} shows calculated intensity ratios for [S II] 6716/6731, [N II] 5755/6583, and [O I] 5577/6300 
within the above temperature ranges
as a function of density. 
We also plot the observed flux ratios for HVC1 and HVC2 with uncertainties including those for the foreground extinction ($A_\mathrm{V} =$ 0.8$\pm$0.8; see Section \ref{sec: discussion LVC}).

For HVC1, we use the [S II] 6716/6731 and [N II] 5755/6583 ratios to estimate the electron densities as these emission lines are not significantly affected by the HVC2 emission (see Section \ref{sec: PV} and Figure \ref{PVDfinal}). 
The left panel of Figure \ref{find hvc density} shows that the observed [S II] 6716/6731 ratio is lower than the calculated values. However, the observed [S II] ratio is very close to the calculated values for $n_e \gtrsim 10^4$ cm$^{-3}$ at or near the LTE conditions. The discrepancies at $n_e \gtrsim 10^4$ cm$^{-3}$ may be due to marginal systematic errors in physical parameters such as the Einstein A coefficients. Therefore, this panel would imply an electron density of HVC1 of $\gtrsim 10^4$ cm$^{-3}$. The middle panel of Figure \ref{find hvc density}, for the [N II] 5755/6583 ratio, indicates an electron density of HVC1 of  $\lesssim 3 \times 10^4$ cm$^{-3}$. Combining these lower and upper limits, these two panels would therefore imply an electron density for HVC1 of $1-3 \times 10^4$ cm$^{-3}$, approximately consistent with those derived using high-angular resolution observations at $\sim$1" away from the DG Tau A star \citep{Lavalley00, Takami23}.\\

For HVC2, we use the [O I] ratio (and therefore the right panel of Figure \ref{find hvc density}) for the emission associated with the region of our interest (see Section \ref{sec: PV} and Figure \ref{HVC2_spatial}). Figure \ref{find hvc density} indicates an electron density of $\sim 10^6$ cm$^{-3}$, significantly higher than that of HVC1.\\

While the [S II] 6716, 6731 Å lines and the [N II] 6583 Å line show HVC2 emission, this is observed at much larger spatial offsets than in the other emission lines (Figure \ref{HVC2_spatial}). Thus, the origin of HVC2 emission in these lines
could differ from the other emission lines (Section \ref{sec: PV}).
Due to the much lower critical densities (Table \ref{tab: line}), we suggest that the HVC2 of these three emission lines 
 probably traces the diffuse gas entrained by the jet knot (HVC1).\\

{
We note that we cannot apply the method to determine the electron temperature described in this section to the LVCs. As discussed in Section \ref{sec: MLR}, the emission lines with lower critical densities show larger positional offsets, indicating the presence of gas at a range of electron densities in the LVCs.
}

\subsection{The low-velocity components (LVCs)}\label{sec: discussion LVC}

As described in Section \ref{sec: major}, we identified up to three LVC components (LVC-H, LVC-M, and LVC-L) in the [O I] 6300 Å, [O I] 5577 Å, and [S II] 6731 Å lines. Of these lines,
the [O I] 5577 Å line profile clearly shows the presence of LVC-M and LVC-L
as well as HVC2,
but not LVC-H.
Compared with [O I] 6300 Å, the [O I] 5577 Å emission is enhanced at high electron densities due to its significantly higher critical density ($1.3 \times 10^8$ and $1.6 \times 10^6$ cm$^{-3}$ for the 5577 Å and 6300 Å lines, respectively; see Table \ref{tab: line}), or at high electron temperatures due to its high upper energy level ($E_u = 3.4 \times 10^4$ and $1.6 \times 10^4$ cm$^{-1}$ for the 5577 Å and 6300 Å lines, respectively; see Table \ref{tab: line}). Therefore, the absence of apparent LVC-H emission in the [O I] 5577 Å indicates that  LVC-H has a lower electron density or temperature than the other components.

In this context, LVC-H may not be associated with regions close to the star or inner disk, 
where
we expect high densities and/or temperatures. 
This can be explained if LVC-H is associated with gas entrained by the fast collimated jet. This interpretation is corroborated by \citet{Pyo03}, who executed near-infrared spectroscopy at high angular resolution ($\sim$0\farcs2) and argued for the detection of a slow entrained component in the [Fe II] 1.64 \micron~emission. In their PV diagram from the observations taken in 2001, the offset of the LVC emission increases with velocity at $|v|=$ --60 to $\sim$--150 km s$^{-1}$ and at 0\farcs25-0\farcs7 from the star. The emission at the highest velocities and spatial offset appeared associated with a jet knot, suggesting that this part of the gas is entrained by the jet.
%
An alternative interpretation of LVC-H 
is a turbulent layer produced by the Kelvin-Helmholtz instability caused by velocity shearing between the jet (HVCs) and the wind layers (LVC-M). This scenario has been studied and proved by a simulation result from
the unified X-wind model proposed by \citet{Shang23} (see Section \ref{sec: X-wind} for details).\\

 The position spectra of LVC-M show a negative velocity gradient from about $-17$ to $0$ km s$^{-1}$ (see Section \ref{sec: major}). Such a gradient was also reported by \citet{Whelan21} for LVC-NC of RU Lupi. These authors explained the gradient by tracing an increase in the height of the wind with increasing disk radius. Therefore, we suggest that LVC-M traces a wide-angled wind as interpreted by \citet{Whelan21}. 
 In Sections \ref{sec: PE or D-wind} and \ref{sec: X-wind}, we will discuss the possibilities for these scenarios in more detail.\\
 
 
LVC-L and the redshifted wing show the highest [O I] 5577/6300 ratio (see Figure \ref{LR profiles}), indicating that they have the highest electron densities or temperatures \citep[Table \ref{tab: line};][]{Gorti11}.
~In addition, the [O I] 5577/6300 line ratio profile shows a 
hump
centered at 0 km s$^{-1}$. 
Therefore, we suggest that LVC-L and the redshifted wing trace the upper disk atmosphere.
%
The above hump is observed over a velocity range $\Delta v$ of $\sim 80$ km s$^{-1}$, suggesting the orbital velocity of the emitting gas of up to $\sim$50 km s$^{-1}$ assuming that the emission is associated with a Keplerian disk with an inclination angle of 35\arcdeg~\citep{Garufi22}. This velocity can be explained if the gas is distributed over radii of $\sim$0.2 au and larger with the given stellar mass of DG Tau A of 0.5 $M_\sun$ \citep{Gudel18}.

One may alternatively attribute the redshifted wing to inflow from the remaining envelope.
Using ALMA, \citet{Garufi22} observed such an inflow 
in the CS $J$=5--4 line
with a velocity towards the disk of $<$ 10 km s$^{-1}$. However, the velocity of shocks induced by this inflow would be too low to produce the observable [O I] 6300 Å and 5577 Å emission, as explained below. 
If all the kinetic energy is converted into thermal energy,
a shock velocity of $\sim$16 km s$^{-1}$ would be required for the shock to heat the gas to $\sim$10$^4$ K, i.e., the temperature required to produce the [O I] 6300 Å and 5577 Å lines, with upper energy levels $E_u/k$=2-5$\times$10$^4$ K (Table \ref{tab: line}).
In practice, more detailed calculations showed that a shock velocity of at least 20-30 km s$^{-1}$ is required to produce such a temperature, as the kinetic energy of the shocks partially escapes through radiative cooling and magnetic precursors \citep[e.g.][]{Hollenbach89, Hartigan94}. We, therefore, rule out the possibility that the redshifted wing is due to an inflow.\\

\begin{figure}
    \centering
    \includegraphics[width=8.5cm]{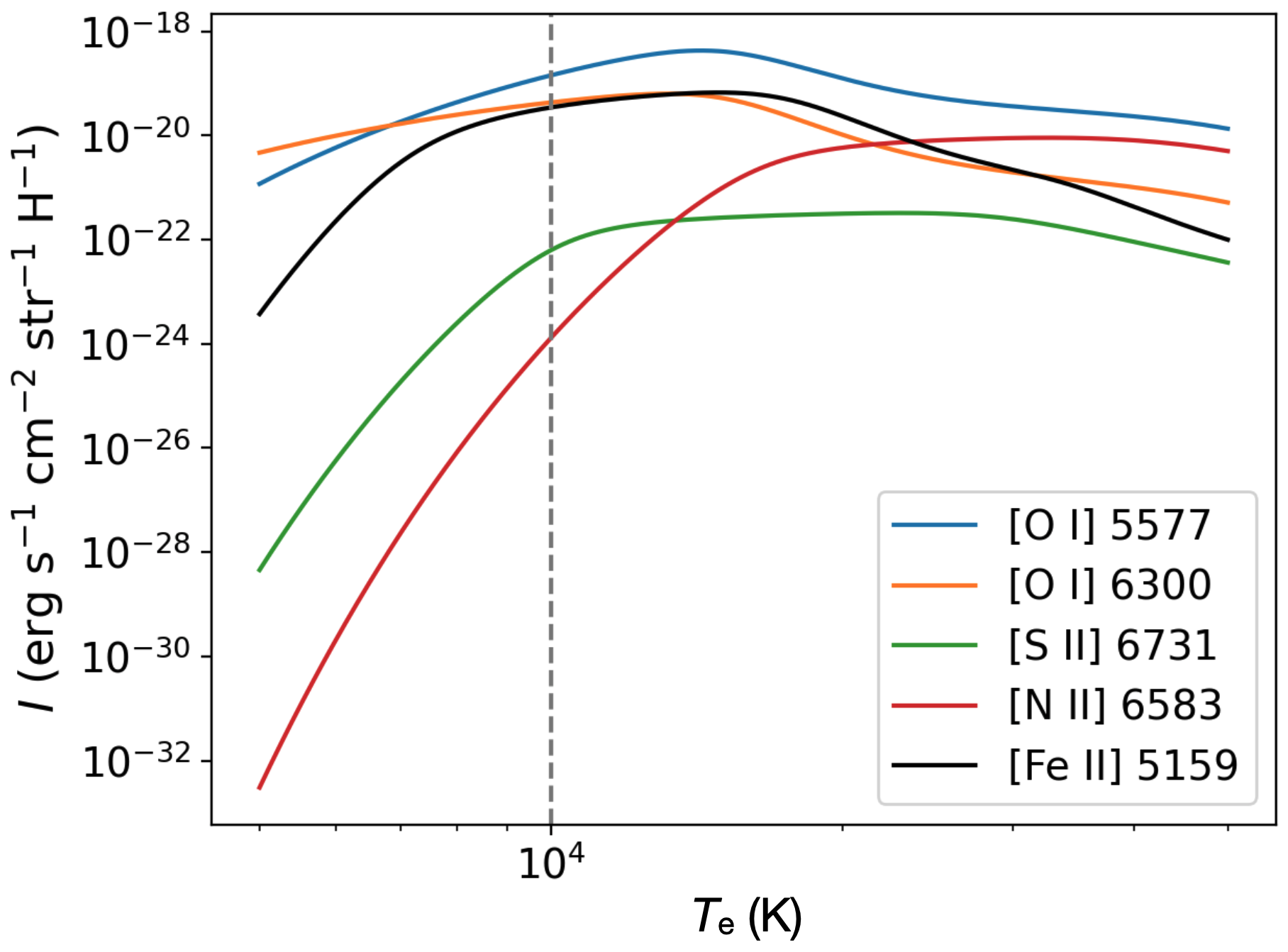}
    \caption{The theoretical intensities of the [O I] 5577, 6300, [S II] 6731, [N II] 6583, and [Fe II] 5159 Å lines as a function of temperature. 
    Each intensity is derived by integrating over the entire frequency or velocity for the line emission. These intensities are scaled by the number of hydrogen atoms.
    That for the [Fe II] 5159 Å line is calculated with Pyneb, while the other lines are calculated with ChiantiPy. The vertical gray dashed line denotes a temperature of $10^4$ K.
    \label{compare intensities}}
\end{figure}

We did not detect LVCs in the iron lines, which indicates iron depletion in the slow wind as explained below.
The [Fe II] 5159 Å line has a similar $n_\mathrm{crit}$ to the [O I] 6300 Å line (Table \ref{tab: line}). 
Furthermore, {the bottom-right panel of} Figure \ref{IvsT_comp} shows that the [Fe II] 5159 Å line and the [O I] 6300 Å line intensities are enhanced at a similar temperature ($\sim1\times10^4$ K). Therefore, the two emission lines share similar emission conditions.
Figure \ref{compare intensities} shows the theoretical intensities of the [O I] 5577, 6300, [S II] 6731, [N II] 6583, and [Fe II] 5159 Å lines as a function of temperatures without any iron depletion.
Each intensity in the figure is scaled by the number of hydrogen atoms to show which emission lines associated with which atoms/ions are brighter or fainter with regard to the given amount of gas.
At $T \sim 10^4$ K, the [Fe II] 5159 Å line intensity is comparable to the [O I] 6300 Å line.
As the LVC is present in the [O I] 6300 Å line, we would have detected low-velocity [Fe II] 5159 Å emission if gaseous iron existed in the wind.
Iron depletion in the LVC(s) of DG Tau A was also reported and discussed by \citet{Giannini19}.\\

To release the iron, the dust must be destroyed by either sublimation or shocks.
For DG Tau A, the dust sublimation radius of the disk is estimated to be located at 0.07 au from the star \citep{Varga17}. According to the X-wind model, the launching radius of the wind is located within this radius, where iron exists in a gas phase \citep{Shu00}. Therefore, the X-wind directly ejected from the disk cannot explain the observed dust depletion of the LVC gas. However, a complex interplay between the X-wind and surrounding gas could still allow the dust grains to exist in the observed gas and, therefore, explain the observations (Section \ref{sec: X-wind}).

\subsection{The Mass-Loss Rates for the LVC Wind} \label{sec: MLR}

\begin{deluxetable}{llll}
\tablewidth{0pt} 
\tablecaption{The parameters used to estimate the wind mass-loss rate \label{tab: wind MLR}}
\tablehead{
\colhead{Line} & \colhead{$v_{\mathrm{p}}$\tablenotemark{\tiny a}  (km s$^{-1}$)} & \colhead{offset\tablenotemark{\tiny b} (au)} & \colhead{$\bm{L_\mathrm{line}}$\tablenotemark{\tiny c} ($10^{22}$ W)}  
} 
\startdata 
[\ion{O}{1}] 5577 & $-25$ & $12$ & $2.9$\\
{[\ion{O}{1}]} 6300 & $-27$ & $28$ & $8.9$\\
{[\ion{S}{2}]} 6731 & $-37$ & $69$ & $5.2$\\
\enddata
\tablenotetext{a}{
The deprojected velocity for the LVC-M wind, assuming the inclination angle of 59\degree~measured for the collimated jet \citep{Takami23}.
}
\tablenotetext{b}{The spatial offset averaged over $-50 \leq v \leq 0$ km s$^{-1}$ from the position spectra (Figure \ref{position}) and corrected for the flow inclination angle ($59$\degree).}
\tablenotetext{c}{The measured line luminosity for the wind.}
\end{deluxetable}

In Section \ref{sec: major} and from Figure \ref{3in1LVC}, we showed the spatial offset of the LVC-M and LVC-L for three emission lines: [O I] 5577 Å, [O I] 6300 Å, and [S II] 6731 Å. As shown in Table \ref{tab: line}, their revised critical densities are 1.3$\times$10$^8$, 1.6$\times$10$^6$ and 6.0$\times$10$^3$ cm$^{-3}$, respectively. The lines with lower critical densities show a larger spatial extent, indicating that the electron density decreases as the distance from the star increases. In this case, the measured offset for each line would approximately correspond to the distance at which its electron density is equal to the critical density. Within this distance, the electron density is higher so that the atom or ion of our interest reaches LTE for the given transition. Beyond this distance, the line intensity decreases as the distance increases, and therefore the electron density decreases.\\

To further investigate the physical nature of the LVC-M, we estimate the mass-loss rates using the individual emission lines under the following assumptions: (1) the region covered by a single emission line is at a single temperature, and (2) their excitation and ionization are dominated by thermal processes. The latter is corroborated by the analysis of \citet{Fang18}, who demonstrated using the [O I] 5577 Å, [O I] 6300 Å, and [S II] 4068 Å lines, that the LVCs are thermally excited in many cases. Furthermore, we approximate that (3) the emission from the region beyond the measured offset is zero for each line, and (4) the regions covered by a single emission line have a single constant velocity along the axis of the wind. These approximations 
should yield estimates accurate to within an order of magnitude, which is sufficient for our discussion below.
As a result, the mass loss rate can be estimated using the following equation, provided by \citet{Fang18}:
\begin{equation} \label{eq:massLossRate}
\dot{M}_\mathrm{wind} = 
\frac{V_\mathrm{wind}}{l_\mathrm{wind}}
\frac{\eta m_\mathrm{H}}{\alpha}
e^{\frac{h \nu}{kT_\mathrm{gas}}}
Z(T_\mathrm{gas})
\frac{L_\mathrm{line}}{g_u~A~h\nu},
\end{equation}
where $V_\mathrm{wind}$ and $l_\mathrm{wind}$ are the measured deprojected velocity and the spectroastrometric offset, respectively;
$\eta$ is the ratio of the total gas mass to the hydrogen mass;
$m_\mathrm{H}$ is the mass of atomic hydrogen;
$\alpha$ is the number fraction of the ion, which is a product of the elemental abundance and the fraction of ionization;
$h \nu$ is the photon energy;
$k$ is the Boltzmann constant;
$T_\mathrm{gas}$ is the gas temperature\footnote{
More precisely, this is the temperature of the atoms/ions determined by the electron populations between the energy levels. Since the calculations are made for LTE conditions (see Section \ref{sec: MLR} for details), this temperature is equal to the electron temperature.
};
$Z(T_\mathrm{gas})$ is the partition function;
$L_\mathrm{line}$ is the line luminosity;
$g_u$ is the statistical weight for the upper level of the transition; $A$ is the Einstein $A$ coefficient (Table \ref{tab: line}).

Table \ref{tab: wind MLR} summarizes the parameters we used to estimate the mass-loss rates. We derived the spatial offsets in the table by averaging the SA offsets measured in the plane of the sky at $v = -$50 to 0 km s$^{-1}$. For 
$V_\mathrm{wind}$, we use the measured deprojected peak velocity 
tabulated in Table \ref{tab: wind MLR}.
We correct these values along the wind axis, adopting the inclination angle of 59\arcdeg~measured for the extended jet
in the [Fe II] 1.644 $\mu$m line
\citep{Takami23}.
\\

To measure $L_\mathrm{line}$, we used the equivalent widths of the LVC-M(+L, which has a minor contribution) measured using the spatial range where the continuum emission is equal to or brighter than 0.3 times its peak (Section \ref{sec: major}).
We adopt a stellar distance of 138 pc, based on the Gaia DR3 measurements \citep{GaiaDR3}, to convert the line fluxes to the luminosities. An extinction of $A_V =$ 1.6 was measured towards the star \citep{Gullbring00}, 
although
for the wind region may be lower. Hence, we corrected the line luminosities for extinction using $A_V =$ 0.8, which is halfway between $A_V =$ 1.6 and $A_V =$ 0. 
We adopted a synthetic extinction curve with $R_V =$ 3.1 \citep{Draine03a, Draine03b} for the dependence of extinction on wavelength. 
As a result, the uncertainty in extinction ($A_V =$ 0-1.6) yields uncertainty in the derived line luminosities of a factor of $\sim$2.
We adopted $\eta =$ 1.41 \citep{Dappen00}, and calculated $\alpha$ (see Appendix \ref{app:alpha}) and exp($h\nu/kT_\mathrm{gas}$) $Z (T_\mathrm{gas}) / (g_u A h\nu)$ in Equation (\ref{eq:massLossRate}) using ChiantiPy. The gas temperature $T_\mathrm{gas}$ is a free parameter in our calculations.\\

 \begin{figure}[ht]
  \includegraphics[width=8.5cm]{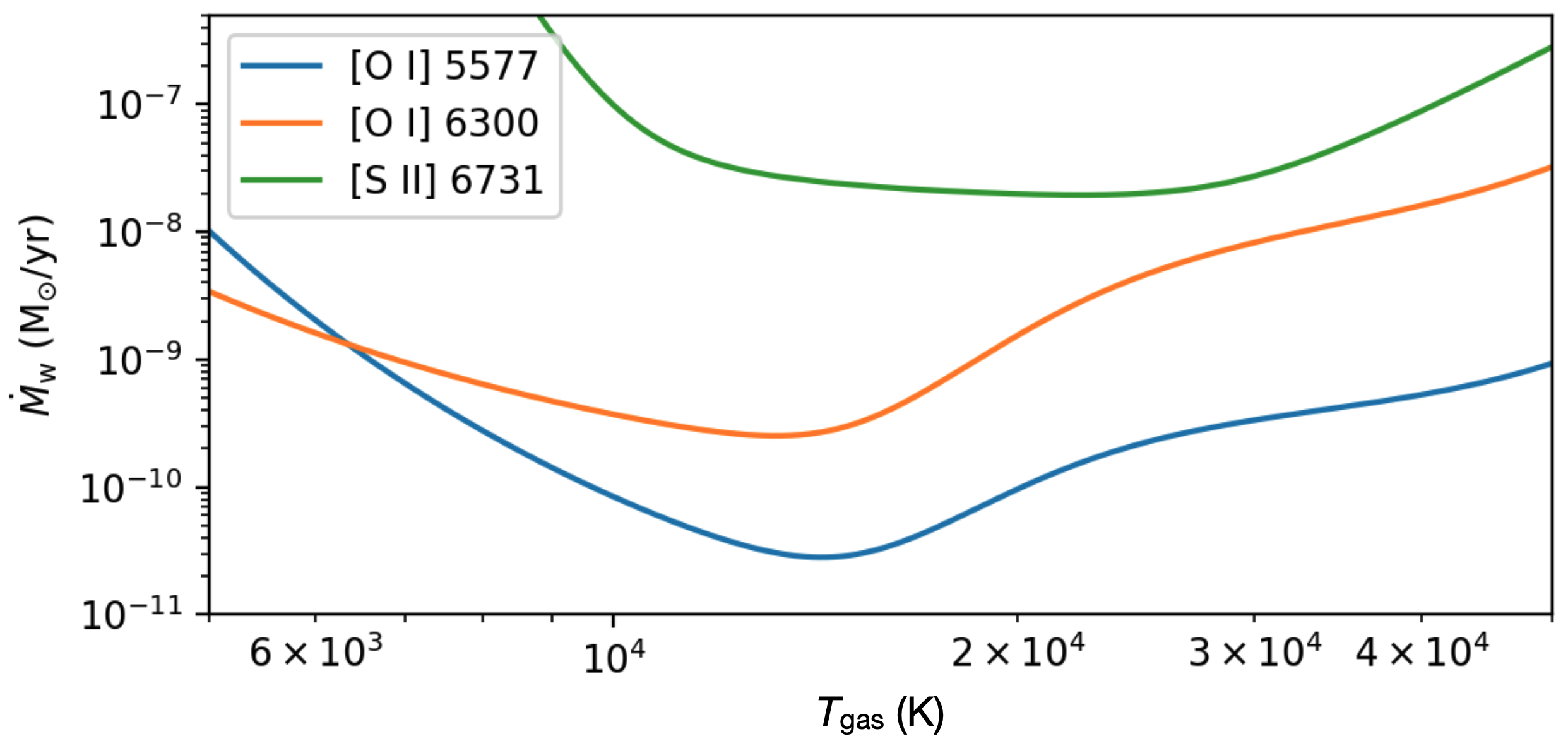}
  \caption{The estimated wind mass-loss rates for [O I] 5577 Å, 6300 Å, and [S II] 6731 Å lines as a function of temperature. See the text for details of the calculations.}
  \label{fig: MLR}
\end{figure}

Figure \ref{fig: MLR} shows the mass-loss rates for the individual emission lines calculated using the above parameters and ChiantiPy.
The figure shows that any single temperature cannot yield a consistent mass-loss rate for the three emission lines. The [S II] 6731 Å line, which covers the wind(s) over the largest length, tends to imply a mass-loss rate larger than those inferred from the [O I] 6300 Å  and 5577 Å lines. The [O I] 6300 Å line, which covers the wind(s) for a larger length than the [O I] 5577 Å line, tends to imply a larger mass-loss rate.\\

The trends described above
can be explained if the emission lines originate from shocked regions and the volume of the shocked regions increases downstream. 
The shocks may be induced either due to (A) internal working surfaces or turbulence in the ejecta, or (B) interaction between the winds and the surrounding gas. For case (A), the fraction of the shock-heated gas in the ejecta would increase with distance from the star. For either case, we derive a minimum mass-loss rate from the disk of 2$\times$10$^{-8}$ $M_\sun$ yr$^{-1}$ based on the mass-loss rates for the [S II] 6731 Å emission shown in Figure \ref{fig: MLR}. 

Alternatively, one might adopt $T_\mathrm{e}$ $\sim$6.3$\times$10$^{3}$ K, which yields a consistent mass-loss rate of $\sim$10$^{-9}$ $M_\odot$ yr$^{-1}$ for the [O I] 6300 Å and 5577 Å lines in Figure \ref{fig: MLR}. However, this temperature would yield a mass-loss rate of $\sim$10$^{-4}$ $M_\odot$ yr$^{-1}$ from the [S II] 6731 Å emission, which is about $10^5$ times larger than that inferred from the above [O I] lines. We do not think such a large discrepancy between the [S II] and [O I] lines is realistic, considering their similar line profiles and spatial scales. 
Furthermore, such a large mass loss rate ($\sim$10$^{-4}$ $M_\odot$ yr$^{-1}$) would yield a dissipation timescale for the circumstellar disk of only a few hundred years with the given disk mass of 3$\times$10$^{-2}$ $M_\sun$ \citep{Varga17}. 
Such a short timescale for dissipation is not realistic considering the stellar age of DG Tau A of $\sim$1 Myr \citep{Muzerolle98_IR,Gullbring00,White04,Dodin20} and a typical lifetime for T Tauri disks of 5-10 Myr \citep{Ribas14}.

\section{
Discussion about the driving mechanism for the LVC-M wind
}\label{sec: mechanism}

In Section \ref{sec: discussion LVC}, we attributed three LVC components (LVC-H, -M, and -L) to the following origins, respectively: (1) gas entrained by the fast collimated jet, or turbulent layers caused by interactions between the jet and the wind; (2) a distinct wind component; and (3) the disk atmosphere. In this section, we discuss the driving mechanism for (2), i.e., LVC-M.
In Section \ref{sec: PE or D-wind}, we compare our results with the PE wind and 
MHD disk wind
models presented by \citet{Weber20}. In Section \ref{sec: X-wind}, we compare our results with the X-wind model proposed by \citet{Shang23}.

\subsection{
Comparisons with the PE 
wind and MHD disk wind models
}\label{sec: PE or D-wind}
\citet{Weber20} modeled emission lines from PE winds and 
MHD disk winds
for different viewing angles (20-40\arcdeg) and accretion luminosities (0.31-1 L$_\odot$), comparable to those for DG Tau A (31\arcdeg~from \citet{Takami23} and 0.56 L$_\odot$ from \citet{Gangi22}.
These authors used PE wind models developed by \citet{Picogna19} and 
MHD disk wind
models developed by \citet{Milliner19}.

We performed comparisons between the models and observations for the line luminosities, velocities, line profiles, and spatial scales. Neither model explains the observations well (Section \ref{sec: PE and D-wind}), even when we combine the PE
wind and MHD disk wind
models (\ref{sec: PE+D-wind}), or when we add another illumination source (\ref{sec: PE+lamppost}) to try to resolve various inconsistencies, as described below.

\subsubsection{PE and 
MHD disk winds
}\label{sec: PE and D-wind}
For all of these models, the predicted luminosities of the [O I] 5577 Å, [O I] 6300 Å, and [S II] 6731 Å lines are dimmer by factors of $\sim$10, $\sim$3, and 2-10 than observed, respectively. Furthermore,
the predicted spatial scales for the [O I] 5577 Å and 6300 Å lines are significantly smaller than the observations. These models yield spatial scales of $<$5 au and $<$1 au for the 6300 Å and 5577 Å lines, 
smaller than the observations by factors of at least 5 and 10, respectively.

Compared with the PE wind models,
the observed line profiles for
these three
lines are much more complex, and
the observed velocities for the LVC winds 
are higher than the models by a factor of $\sim$5. 
The 
MHD disk wind
models yield velocities of 20-40 km s$^{-1}$, comparable to the observations. 
However,
the modeled LVC line profiles
exhibit two peaks
associated with the two sides of 
a wind that is spiraling upwards from the disk. Similar peaks
are not 
seen 
in our [O I] 6300 Å profile, or those observed by other authors \cite[e.g.,][see also Weber et al. 2020]{Banzatti19}.

%
%
%
%
%
%

\subsubsection{
Hybrid Models for PE and MHD disk winds
}\label{sec: PE+D-wind}

\citet{Weber20} combined the modeled [O I] 6300 Å line profiles for the PE 
wind and MHD disk wind
models to attempt to solve the inconsistency of the double-peak profile not being observed in LVCs for DG Tau A (Section \ref{sec: PE and D-wind}) and many other T-Tauri stars \citep[e.g.,][]{Banzatti19}.
The combined line profile ($i =$ 40\arcdeg, $L_\mathrm{acc} =$ 0.31 L$_\odot$) qualitatively reproduces the peak for LVC-M as well as a hump similar to LVC-H.
However, the flux ratio of this hump to the LVC is $\sim$0.5 in the modeled line profile, lower than our observations ($\sim$0.8; Figure \ref{3in1LVC}).
This difference may be attributed to different inclination angles or accretion luminosities between the models and the observations.\\ 

However, the predicted peak velocities are similar to those for the PE winds, and as a result, the modeled emission is still less blueshifted than the observations (Section \ref{sec: PE and D-wind}).

\subsubsection{PE wind with Lamppost Illumination}\label{sec: PE+lamppost}
The ``less-blueshifted'' problem mentioned in Sections \ref{sec: PE and D-wind} and \ref{sec: PE+D-wind} may be resolved by ``lamppost illumination'' by a shock in the extended jet.
\citet{Weber20} showed that the radiation heating from a shock in a jet, modeled as a blackbody at 25000 K and 50 au above the midplane, can enhance the line emission from flow components at higher velocities and, therefore, yield a larger blueshift in the line profiles  (see their Figure 12). 
The authors also showed that the heating by a shock creates an additional emitting region for the [S II] 6731 and [O I] 6300 Å lines but not the [O I] 5577 Å line (c.f. their Figure 11).\\

However, according to the simulated spatial distribution of the [O I] 6300 and [S II] 6731 Å lines, those two lines should have similar spatial offsets above the midplane.
In contrast, our observations show that the [S II] 6731 Å line has a spatial offset two times larger than that of the [O I] 6300 Å line (see Table \ref{tab: wind MLR}).
Also, the issue of the small spatial distribution of the [O I] 5577 Å line in the PE wind model still exists in this combined model.
Therefore, the PE wind with lamppost illumination model cannot explain the observed wind from DG Tau A very well. 
It would be insightful to further investigate how the shock heating will affect the line profiles in the 
MHD disk wind
(and also the X-wind; see Section \ref{sec: X-wind}) models, and how well those synthetic observations match our results.\\

\subsection{Comparisons with the X-wind model}\label{sec: X-wind}

\citet{Shang23} carried out extensive numerical simulations using the X-wind models, which yield complex outflow structures via interactions with the surrounding gas. According to their simulations, the wind and the jet grow together from the innermost region of the disk in the form of an elongated non-spherical outflow bubble with multi-cavities formed by the jet-wind and wind-ambient medium interactions.
The outflow structures consist of a collimated jet, a magnetized free wind without interactions from the launching point, a reverse shock, a compressed wind,
a wind-ambient interface (tangential discontinuity), a compressed ambient medium, and a forward shock, from the inner to the outer radii.
The reverse shock and forward shock formed in their simulations are fast shocks, which increase the angle between the shock normal and the plasma flow direction in the postshock region \citep[cf. Figure 21 in][]{Shang20}. The authors argue that this deflection causes an optical illusion that makes a
low-velocity
wind appear to be launched from an outer radius of the disk.\\


The observed three LVC components may be explained by the X-wind model as follows.
LVC-H and LVC-M may be attributed to gas between the reverse shocks and the compressed ambient gas at remarkably 
different spatial offsets. The LVC-H would be associated with the turbulence induced by pseudopulses and Kelvin-Helmholtz instabilities for the extended collimated jet. In contrast, the LVC-M would be associated with layers of gas significantly closer to the wind base at higher electron densities. More detailed investigation requires synthetic observations of the emission lines modeled for the individual components predicted by \citet{Shang23}.\\


\section{Summary and Conclusions} \label{sec: summary}

We analyzed high-spectral resolution
observations
($\Delta v$ $\sim$ 2.5 km s$^{-1}$) 
of
DG Tau A and identified 26 emission lines from 4800 Å to 7500 Å, including H$\alpha$, H$\beta$, and He I, associated with the jet and winds. We identified two HVCs (HVC1 and HVC2 at $v$ $\sim$ --250 and --185 km s$^{-1}$, respectively) in these emission lines. Three LVCs
(LVC-H, LVC-M, and LVC-L 
identified at $\sim$--70, $\sim$--30 and $\sim$--10 km s$^{-1}$, respectively) 
and a redshifted wing were found in the [O I] 6300 Å line profile. Some of the LVC components were also observed in the [O I] 5577 Å and [S II] 6731 Å lines.
The LVCs were not clearly identified in the other lines due to either low signal-to-noise, residual subtraction of the stellar continuum, or contaminating emission from the other lines.
\\

We applied spectroastrometry to obtain the position spectra of the emission lines following \citet{Takami01}.
The offset position spectra of the [SII] 6731 Å line cover an extended emission region above 
the disk, with a larger spatial extent than the [OI] 6300 Å and 5577 Å line emissions.
The [O I] 5577 Å line is an outflow-base tracer due to its small positional displacements from the central star.
The positional offset increases with increasing blueshifted velocity ($v <$ --30 km s$^{-1}$) for the [O I] 6300 Å and [S II] 6731 Å lines. However, the opposite trend (the ``negative velocity gradient") was observed at less blueshifted velocities ($v > -30$ km s$^{-1}$).\\

We summarize our interpretations below:
\begin{enumerate}
    \item 
    The HVCs originate from the postshock regions. This interpretation is corroborated by the fact that these components are observed in emission lines from significantly different ionization conditions.
    The HVC1, which has a lower density ($n_e$ $\sim$10$^4$ cm$^{-3}$), is associated with an internal shock surface $\sim$1” away from the central star. The HVC2, which has a higher density ($n_e$ $\sim$10$^6$ cm$^{-3}$), is located at the jet base ($\sim$0\farcs2), and may be a stationary shock component as suggested by some previous studies.
    \item 
    We suggest that the LVC-H traces entrained gas or a turbulent layer. The LVC-H is absent in the [O I] 5577 Å line but evident in the [O I] 6300 Å and [S II] 6731 Å lines. As the [O I] 5577 Å line traces the outflow base, the LVC-H seems not to originate from the disk.

    \item 
    The negative velocity gradient observed in the [O I] 6300 Å, [O I] 5577 Å, and [S II] 6731 Å can be attributed to the presence of a wide-angled wind, as originally proposed by \citet{Whelan21}. We attribute this feature in the position spectra to LVC-M, suggesting that LVC-M is associated with a wide-angled wind.
    
    \item 
    The LVC-L and the redshifted wing originate from the upper disk atmosphere.
    This interpretation is inferred from the high [O I] 5577/6300 intensity ratios observed at $-$50 to 50 km s$^{-1}$, centering at zero velocity and covering these two emission components. The observed line width indicates the innermost radius of the disk atmosphere is about 0.3 au.
    
    \item 
    Using spectroastrometry, we measured deprojected lengths 
    for LVC-M of 12, 28, and 69 au in the [O I] 5577 Å, [O I] 6300 Å, and [S II] 6731 Å lines, respectively. The length increases as the critical density of the emission lines decreases.
    The inferred mass loss rate also increases as the length measured in the emission line increases, probably because of an increasing volume of the shocked gas. From these results, we estimate a lower limit wind mass loss rate of $\sim$10$^{-8}$ M$_\odot$ yr$^{-1}$.
    
    \item 
    As previously reported by \citet{Giannini19}, the [Fe II] lines do not show the presence of LVCs, indicating that iron is depleted from dust grains in these emission components. This implies that 
    the gas emission comes from outside the dust sublimation radius.
    Alternatively, the outflowing gas could be ambient gas interacting with either the X-wind or 
    the MHD disk wind
    without experiencing heating over the dust sublimation temperature.
    
    \item 
    Neither the present PE wind nor the 
    MHD disk wind
    model used in \citet{Weber20} can solely explain the observed 
    lines profiles and spatial scales of the LVCs.
    Even so, we do not exclude the possibility that more complex physical conditions not explored by \citet{Weber20} (e.g., a combination of the PE wind and the 
    MHD disk winds
    with lamppost illumination) could explain our observations.
    The above observations may be alternatively explained by the X-wind model, which predicts complex nested kinetic structures in the ejected gas and interactions with surrounding gas. Synthetic observations of the emission lines for these features are needed to test the theories further.

\end{enumerate}

Direct confirmation of the nature of LVC-H, which is located at 0\farcs2--0\farcs4 from the central star, may be possible using the Very Large Telescope (VLT) or Subaru Telescope with adaptive optics (AO) at optical wavelengths. Both telescopes have a diffraction-limited angular resolution of about 0\farcs02 at the target wavelength ($\sim$6300 Å).
It may also be possible to confirm the nature of LVC-L and the redshifted wing as components of the disk atmosphere if we obtain position-position-velocity data cubes in the [O I] 6300 Å and 5577 Å lines with the VLT Multi-Unit Spectroscopic Explorer (MUSE) or Subaru Faint Object Camera and Spectrograph (FOCAS). 
Instruments with improved spectral resolution may be needed if we have to spectrally resolve the LVC-H/-M/-L and HVC, as the velocity resolutions of MUSE and FOCAS are around 100--130 km s$^{-1}$ at the target wavelength.
The most significant remaining puzzle in our work is distinguishing between the X-wind and 
MHD disk wind
models through observation. A synthetic observation for the X-wind model that provides line profiles, offset position spectra and line luminosities would be useful for further discussion.

\begin{acknowledgments}
We are grateful to Drs. A. Chrysostomou, F. Bacciotti, C. Dougados, and S. Cabrit, and Professor T.P. Ray for their assistance in obtaining the observing time at Subaru.
M.T. and Y.-R.C. are supported by the National Science and Technology Council (NSTC) of Taiwan (grant No. 110-2112-M-001-044-, 111-2112-M-001-002-, 112-2112-M-001-031-, 113-2112-M-001-009-).
S.-P.L.  acknowledges the grant from the NSTC of Taiwan 109-2112-M-007-010-MY3 and 112-2112-M-007 -011.
H.S. is supported by the NSTC grant 112-2112-M-001-030- {and 113-2112-M-001-008-}.
This work has made use of data from the European Space Agency (ESA) mission Gaia (https://www.cosmos.esa.int/gaia), processed by the Gaia Data Processing and Analysis Consortium (DPAC, https://www.cosmos.esa.int/web/gaia/dpac/consortium). Funding for the DPAC has been provided by national institutions, in particular the institutions participating in the Gaia Multilateral Agreement.
This research made use of the Simbad database operated at CDS, Strasbourg, France, and the NASA's Astrophysics Data System Abstract Service.
The authors wish to recognize and acknowledge the very significant cultural role and reverence that the summit of Mauna Kea has always had within the indigenous Hawaiian community.  We are most fortunate to have the opportunity to conduct observations from this mountain.
\end{acknowledgments}

\facility{Subaru (HDS)}

\appendix

\section{ Revised vs. Conventional Critical Densities}\label{app:n_crit}

{
Figure \ref{fig:n_crit_rev} shows the intensities normalized to the LTE values ($I/I_\mathrm{LTE}$) as a function of electron density and for electron temperatures $T_e$ of 3$\times$10$^3$, 1$\times$10$^4$, and 3$\times$10$^4$ K.
These were calculated using ChiantiPy and Pyneb as described in Section \ref{sec: discussion calc} in detail.
For the [Fe III] 5527 \AA~line, the results are shown only for $T_e$=1$\times$10$^4$ and 3$\times$10$^4$ K as the the temperature of 3$\times$10$^3$ K is too low for the reliable calculations using ChiantiPy due to its very high upper level energy (see Table \ref{tab: line}). The calculations were made for electron densities $n_e$ up to 10$^{12}$ cm$^{-3}$, which is sufficient for determining the LTE values except for the [Fe II] 5527 \AA~ line. Each panel shows that the revised critical density, which corresponds to $I/I_\mathrm{LTE}$=0.5, has a weak dependence on the electron temperature.

\begin{figure*}
    \centering
    \includegraphics[width=1.0\textwidth]{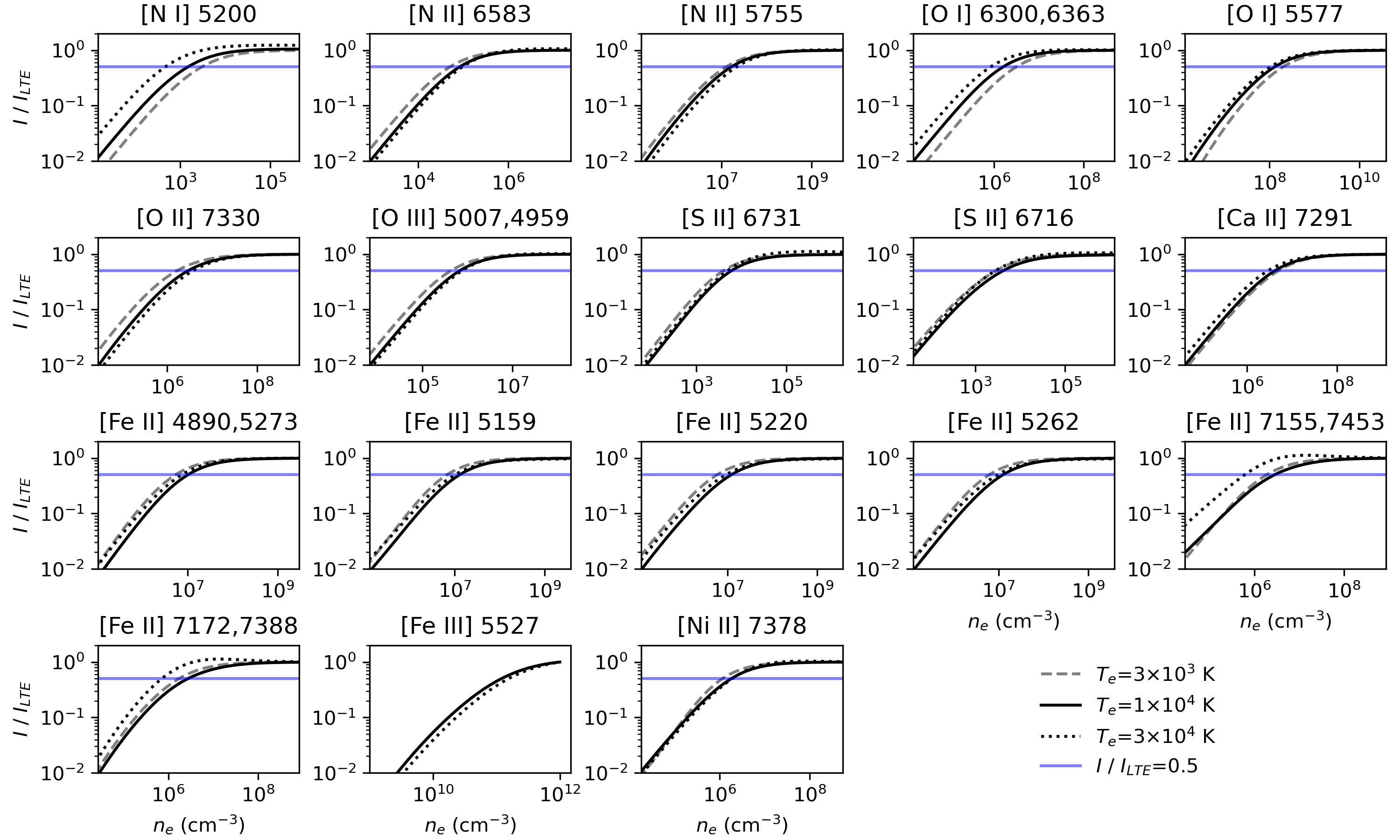}
    \caption{ The intensities normalized to the LTE values ($I/I_\mathrm{LTE}$) as a function of electron density and for the electron temperatures of 3$\times$10$^3$, 1$\times$10$^4$, and 3$\times$10$^4$ K. The [O I] 6300 and 6363 \AA, [O III] 5007 and 4959 \AA, [Fe II] 4890 and 5273 \AA, [Fe II] 7155 and 7453 \AA, and  [Fe II] 7172 and 7388 \AA~lines yield identical results due to the same upper energy level. The blue horizontal line in each panel shows $I/I_\mathrm{LTE}$=0.5, used to determine the revised critical densities.
    This line is not shown for the [Fe II] 5527 \AA~ line, for which the range of electron densities used for the calculations is not sufficient to determine the LTE intensity.
    }
    \label{fig:n_crit_rev}
\end{figure*}

In Table \ref{tbl:n_crit}, we tabulated the revised critical densities for the above three electron temperatures with the conventional critical densities calculated using the ready-made function of Pyneb. 
For the majority of the transitions, the discrepancies between the conventional and revised critical densities are within $\pm$50 \%, therefore these are approximately consistent. The use of $I/I_\mathrm{LTE}$=0.3 or 0.7 instead of $I/I_\mathrm{LTE}$=0.5 would alter the revised critical densities by a factor of $\sim$2, yielding systematic discrepancies from the conventional critical densities. 
The scaling factor of 0.5 was originally provided by \citet{Takami10b} based on the fact that 
the equation is approximately correct at the conventional critical density for the two-level model. More detailed adjustment of this scaling factor is beyond the scope of the paper.

The conventional critical densities are significantly lower than the revised critical densities (by a factor of $>$2) for the other transitions. See Section \ref{sec: discussion calc} for the reason. As described in Section \ref{sec: discussion calc}, the critical densities are used to concisely discuss electron densities for emission line regions, as in many cases the emission is enhanced above the critical density at which the gas reaches LTE. For this purpose, the revised critical densities should be more appropriate than the conventional critical densities as shown in Figure \ref{fig:n_crit_rev}.
}

\section{Line Intensities as a Function of Electron Temperature}\label{app:f8}

In Figure \ref{IvsT_comp}, we plotted the calculated intensities of a subset of emission lines as a function of electron temperature. As described in Section \ref{sec: discussion HVC}, 
the intensity (and therefore the emissivity) for
each line is
high
over a specific temperature range.
This range is nearly independent over a 10$^4$ range in electron density. In this section, we discuss the causes of these trends.

\begin{figure}
    \centering
    \includegraphics[width=1.0\textwidth]{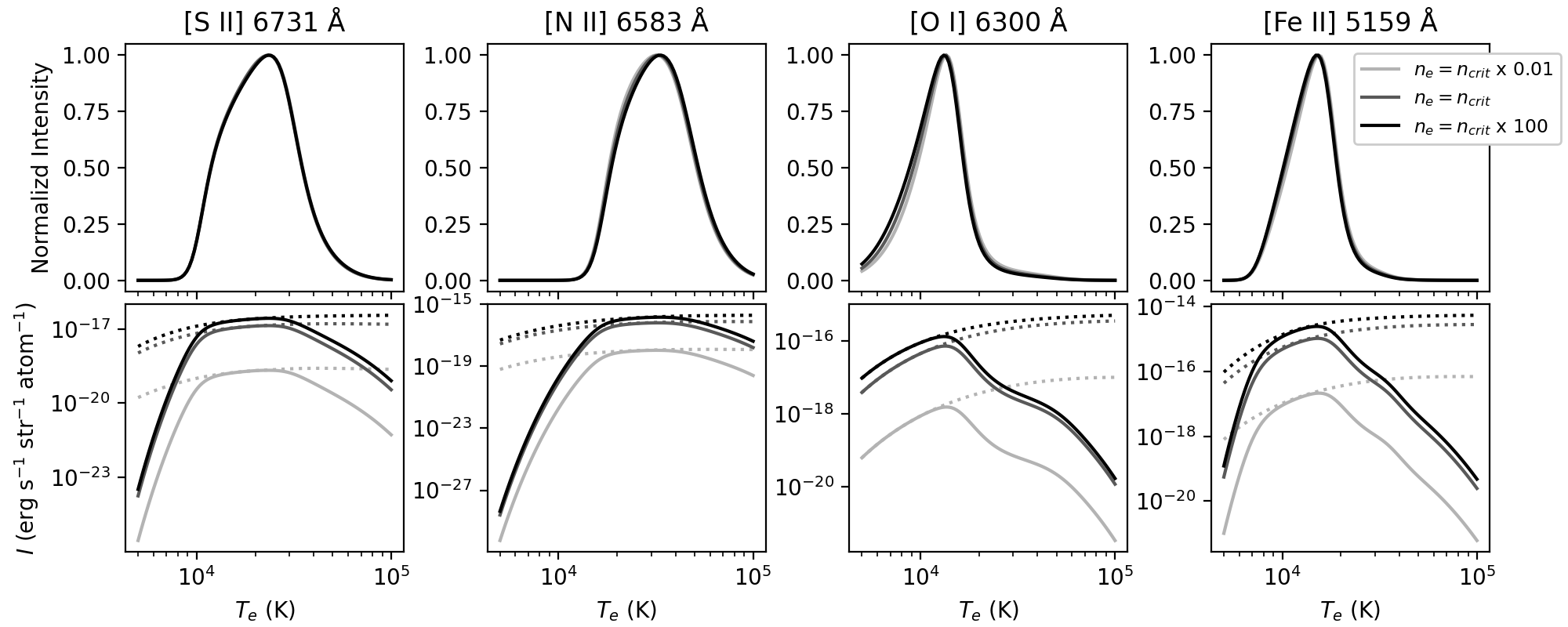}
    \caption{\textit{Upper}: Same as Figure \ref{IvsT_comp}. As in Figure \ref{IvsT_comp}, each curve is normalized to the peak intensity. \textit{Lower}: The intensity for the same cases. The solid curves are identical to those in the upper panels, considering the ionization fraction, but before normalizing by the peak intensity. The dotted curves show the cases which assume the ionization fraction is unity. Therefore, the deviation of the solid curves from the dotted curves indicates the ionization fraction.}
    \label{to discuss Fig8}
\end{figure}

In the upper panels of Figure \ref{to discuss Fig8}, we show a subset of the curves in Figure \ref{IvsT_comp}. In the bottom panels, we show the calculated intensities for some lines as a function of electron temperature. These are shown in the unit of erg s$^{-1}$ str$^{-1}$ atom$^{-1}$, i.e., per element particle including all the ionization states. The solid curves in these panels show the intensities considering the ionization fraction. These curves show that the intensity scales with the electron density, but without significantly changing the dependence on temperature. This trend results in nearly the same curves in the upper panels after normalizing each of them by the peak value.

To further discuss the causes of this trend, we separate the following two physical processes that determine the line intensity: (1) the ionization fraction of the relevant atoms or ions for the emission line; and (2) the dependence of the intensity on electron temperatures and densities per atom/ion  (hereafter ``line excitation''). 
The former is independent of electron density as both the ionization and recombination rates are proportional to the electron density. To further investigate the line excitation, we plot intensities setting the ionization fraction to unity, shown with dotted curves in the bottom panels of Figure \ref{to discuss Fig8}. 
As for the solid curves, the dashed curves also scale with the electron density but without significantly changing the dependence on temperature. Therefore, the solid curves show the same trend regardless of the temperature ranges at both high and low ionization fractions, as indicated by the deviation of the solid curves from the dotted curves. 
Again, this results in nearly the same curves in the upper panels after normalizing each of them by the peak value. See \citet{Osterbrock89} for details of the individual physical processes and their trends described in this section.

\section{Ionization Fraction as a Function of Temperature}\label{app:alpha}

\begin{figure}
    \centering
    \includegraphics[width=0.5\textwidth]{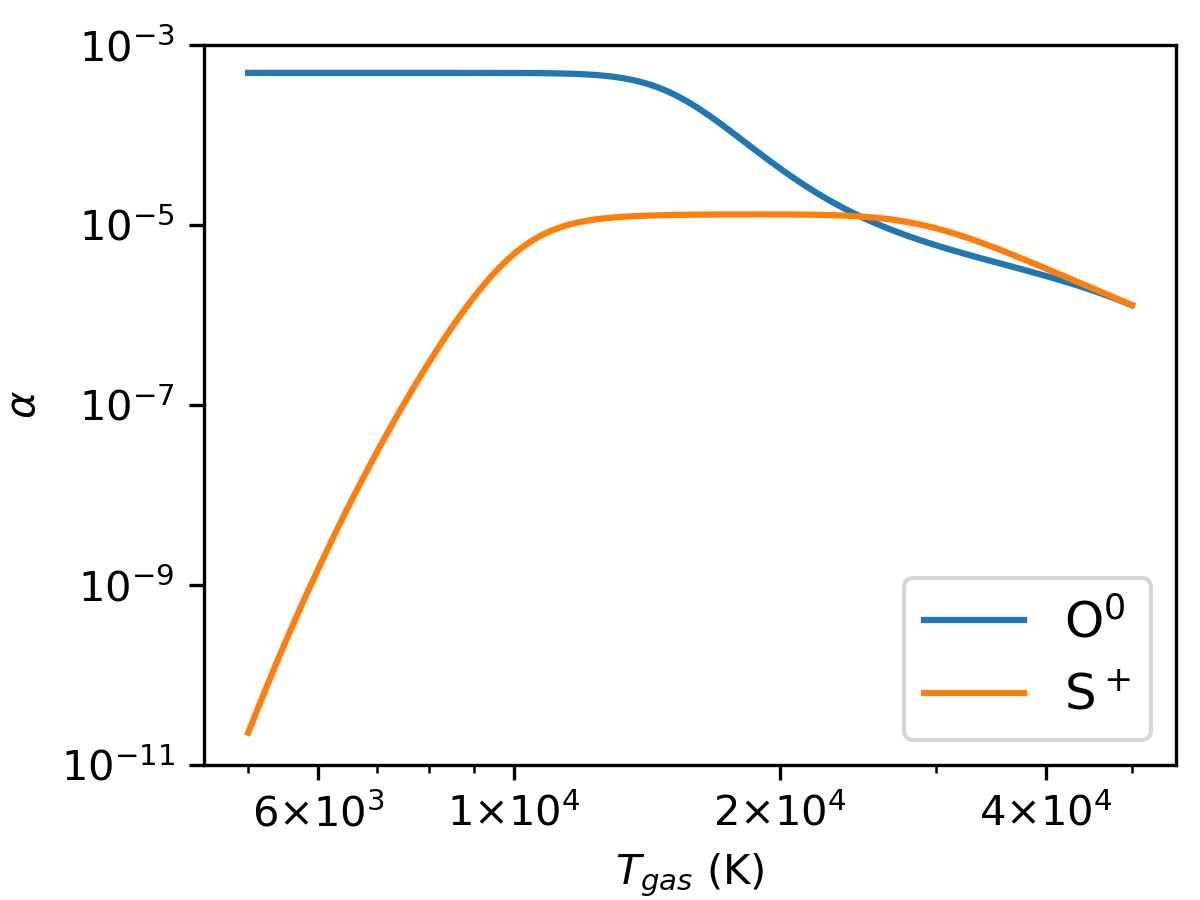}
    \caption{The parameter $\alpha$ for Equation (\ref{eq:massLossRate}) calculated using ChiantiPy. See Appendix \ref{app:alpha} and Section \ref{sec: MLR} for details.}
    \label{fig:alpha}
\end{figure}

Figure \ref{fig:alpha} shows the parameter $\alpha$ (i.e. the number fraction of the ion, which is a product of the elemental abundance and the fraction of ionization) for Equation (\ref{eq:massLossRate}), for the O$^0$ atoms and the S$^+$ ions, as a function of gas temperature. These were calculated using ChiantiPy (see Section \ref{sec: discussion calc}).
{
The value for the O$^0$ atoms corresponds to an elemental abundance of 4.8$\times$10$^{-4}$ at $T_\mathrm{gas}$$\lesssim$1.2$\times$10$^4$ K due to the ionization fraction of $\sim$1, and it decreases at higher temperatures due to the ionization of the oxygen atoms. 
Similarly, the value for the S$^+$ ions corresponds to an elemental abundance of 1.3$\times$10$^{-5}$ at $T_\mathrm{gas}$$\sim$1.5-2.5$\times$10$^4$ K, and it decreases at lower and  higher temperatures due to lower and higher ionization, respectively. 
}

\begin{deluxetable*}{lllccccccccccc}
\tabletypesize{\scriptsize}
\tablecaption{ Conventional vs. Revised Critical Densities  \label{tbl:n_crit}}
\tablehead{
\multicolumn{2}{c}{ Transition} &&
\multicolumn{3}{c}{ $T_e$=3$\times$10$^3$ K}
&&
\multicolumn{3}{c}{ $T_e$=1$\times$10$^4$ K}
&&
\multicolumn{3}{c}{ $T_e$=3$\times$10$^4$ K}
\\
\cline{4-6}
\cline{8-10}
\cline{12-14}
&
&& \colhead{ (conv)\tablenotemark{\tiny a}}
& \colhead{ (rev)\tablenotemark{\tiny b}}
& \colhead{ (conv/rev)\tablenotemark{\tiny c}}
&& \colhead{ (conv)\tablenotemark{\tiny a}}
& \colhead{ (rev)\tablenotemark{\tiny b}}
& \colhead{ (conv/rev)\tablenotemark{\tiny c}}
&& \colhead{ (conv)\tablenotemark{\tiny a}}
& \colhead{ (rev)\tablenotemark{\tiny b}}
& \colhead{ (conv/rev)\tablenotemark{\tiny c}}
}
\startdata
$[$N I$]$ &5200        &&9.2$\times$10$^{2}$ & 3.0$\times$10$^{3}$ & 0.30 &&7.0$\times$10$^{2}$ & 1.5$\times$10$^{3}$ & 0.47 &&5.6$\times$10$^{2}$ & 4.6$\times$10$^{2}$ & 1.21\\
$[$N II$]$ &6583       &&5.0$\times$10$^{4}$ & 5.1$\times$10$^{4}$ & 1.00 &&8.9$\times$10$^{4}$ & 8.4$\times$10$^{4}$ & 1.06 &&1.3$\times$10$^{5}$ & 9.6$\times$10$^{4}$ & 1.40\\
$[$N II$]$ &5755       &&1.2$\times$10$^{7}$ & 1.3$\times$10$^{7}$ & 0.98 &&1.6$\times$10$^{7}$ & 1.7$\times$10$^{7}$ & 0.99 &&2.4$\times$10$^{7}$ & 2.2$\times$10$^{7}$ & 1.12\\
$[$O I$]$ &6300, 6363   &&2.0$\times$10$^{6}$ & 3.1$\times$10$^{6}$ & 0.65 &&1.5$\times$10$^{6}$ & 1.6$\times$10$^{6}$ & 0.96 &&9.8$\times$10$^{5}$ & 8.4$\times$10$^{5}$ & 1.17\\
$[$O I$]$ &5577        &&8.8$\times$10$^{7}$ & 2.0$\times$10$^{8}$ & 0.44 &&9.5$\times$10$^{7}$ & 1.3$\times$10$^{8}$ & 0.73 &&8.9$\times$10$^{7}$ & 1.0$\times$10$^{8}$ & 0.85\\
$[$O II$]$ &7330       &&2.3$\times$10$^{6}$ & 1.6$\times$10$^{6}$ & 1.39 &&4.0$\times$10$^{6}$ & 2.9$\times$10$^{6}$ & 1.38 &&6.0$\times$10$^{6}$ & 3.7$\times$10$^{6}$ & 1.61\\
$[$O III$]$ &5007,4959 &&4.3$\times$10$^{5}$ & 4.3$\times$10$^{5}$ & 0.99 &&6.9$\times$10$^{5}$ & 6.8$\times$10$^{5}$ & 1.02 &&1.0$\times$10$^{6}$ & 7.6$\times$10$^{5}$ & 1.32\\
$[$S II$]$ &6731       &&2.4$\times$10$^{3}$ & 4.0$\times$10$^{3}$ & 0.61 &&4.2$\times$10$^{3}$ & 6.0$\times$10$^{3}$ & 0.71 &&7.0$\times$10$^{3}$ & 4.9$\times$10$^{3}$ & 1.45\\
$[$S II$]$ &6716       &&9.5$\times$10$^{2}$ & 2.9$\times$10$^{3}$ & 0.33 &&1.6$\times$10$^{3}$ & 4.2$\times$10$^{3}$ & 0.39 &&2.6$\times$10$^{3}$ & 2.8$\times$10$^{3}$ & 0.93\\
$[$Ca II$]$ &7291      &&---\tablenotemark{ \tiny d} & 4.9$\times$10$^{6}$ & ---\tablenotemark{ \tiny d} &&---\tablenotemark{ \tiny d} & 4.1$\times$10$^{6}$ & ---\tablenotemark{ \tiny d} &&---\tablenotemark{ \tiny d} & 2.9$\times$10$^{6}$ & ---\tablenotemark{ \tiny d}\\
$[$Fe II$]$ &4890,5273 &&3.3$\times$10$^{6}$ & 5.4$\times$10$^{6}$ & 0.60 &&4.5$\times$10$^{6}$ & 1.0$\times$10$^{7}$ & 0.45 &&4.5$\times$10$^{6}$ & 7.4$\times$10$^{6}$ & 0.61\\
$[$Fe II$]$ &5159      &&2.3$\times$10$^{6}$ & 6.8$\times$10$^{6}$ & 0.33 &&2.7$\times$10$^{6}$ & 1.3$\times$10$^{7}$ & 0.21 &&2.3$\times$10$^{6}$ & 1.0$\times$10$^{7}$ & 0.23\\
$[$Fe II$]$ &5220      &&6.9$\times$10$^{5}$ & 5.9$\times$10$^{6}$ & 0.12 &&8.4$\times$10$^{5}$ & 1.2$\times$10$^{7}$ & 0.07 &&9.5$\times$10$^{5}$ & 9.3$\times$10$^{6}$ & 0.10\\
$[$Fe II$]$ &5262      &&1.1$\times$10$^{6}$ & 6.2$\times$10$^{6}$ & 0.17 &&1.2$\times$10$^{6}$ & 1.3$\times$10$^{7}$ & 0.09 &&1.3$\times$10$^{6}$ & 9.5$\times$10$^{6}$ & 0.13\\
$[$Fe II$]$ &7155,7453 &&6.7$\times$10$^{5}$ & 1.9$\times$10$^{6}$ & 0.35 &&5.8$\times$10$^{5}$ & 2.8$\times$10$^{6}$ & 0.20 &&6.0$\times$10$^{5}$ & 6.1$\times$10$^{5}$ & 0.98\\
$[$Fe II$]$ &7172,7388 &&3.5$\times$10$^{5}$ & 1.7$\times$10$^{6}$ & 0.20 &&5.1$\times$10$^{5}$ & 2.7$\times$10$^{6}$ & 0.19 &&6.0$\times$10$^{5}$ & 7.2$\times$10$^{5}$ & 0.84\\
$[$Fe III$]$ &5527     &&1.8$\times$10$^{8}$ & ---\tablenotemark{ \tiny e} & ---\tablenotemark{ \tiny e} &&3.0$\times$10$^{8}$ & >1$\times$10$^{12}$ & <10$^{-3}$ &&5.3$\times$10$^{8}$ & >1$\times$10$^{12}$ & <10$^{-3}$\\
$[$Ni II$]$ &7378      &&---\tablenotemark{ \tiny d} & 1.2$\times$10$^{6}$ & ---\tablenotemark{ \tiny d} &&---\tablenotemark{ \tiny d} & 1.8$\times$10$^{6}$ & ---\tablenotemark{ \tiny d} &&---\tablenotemark{ \tiny d} & 1.9$\times$10$^{6}$ & ---\tablenotemark{ \tiny d}
\enddata
\tabletypesize{\normalsize}
\tablenotetext{ a}{ Conventional critical density}
\tablenotetext{ b}{ Revised critical density}
\tablenotetext{ c}{ Ratio of the conventional critical density per revised critical density}
\tablenotetext{ d}{ The Pyneb calculations for the critical density is not available because the transition is not included.}
\tablenotetext{ e}{ The temperature is too low for the CHIANTI calculations.}
\end{deluxetable*}

\bibdata{astro.bib}
\bibliography{astro}
\bibliographystyle{aasjournal}

\end{document}